\documentclass[linenumbers]{aastex}
\usepackage{amsmath}
\usepackage{comment}
\usepackage{grffile}
\usepackage{lscape}
\usepackage{arrayjobx}
\usepackage{multirow}

\defcitealias{Oya_16293Cyc4}{OY20} 
\bibpunct[]{(}{)}{;}{a}{}{,}

\newcommand{\bff}{}

\newcommand{\CO}{C$^{17}$O}
\newcommand{\co}{$J$ = 2--1}
\newcommand{\CS}{C$^{34}$S}
\newcommand{\cs}{$J$ = 2--1}
\newcommand{\so}{$J_N$ = $2_2$--$1_1$}
\newcommand{\ocsa}{$J$ = 7--6}

\newcommand{\sio}{$J$ = 2--1}
\newcommand{\TFA}{H$_2$CS}

\newcommand{\MN}{CH$_3$OH}
\newcommand{\mn}{$5_{3,2}$--$6_{2,4}$}

\newcommand{\MF}{HCOOCH$_3$}
\newcommand{\DME}{(CH$_3$)$_2$O}
\newcommand{\PPN}{C$_2$H$_5$CN}

\newcommand{\coYeha}{$J$ = 2--1}
\newcommand{\coYehb}{$J$ = 3--2}
\newcommand{\coRao}{$J$ = 3--2}
\newcommand{\coKristensen}{$J$ = 6--5}
\newcommand{\coGirart}{$J$ = 3--2}
\newcommand{\coLoinard}{$J$ = 6--5}

\newcommand{\iras}{IRAS 16293--2422}

\newcommand{\ire}{infalling-rotating envelope}

\newcommand{\pos}{plane of the sky} 
\newcommand{\cb}{centrifugal barrier} 

\newcommand{\desys}{disk/envelope system}

\newcommand{\ia}{inclination angle}

\newcommand{\incRemOutflow}{0\degr\ for a pole-on configuration}
\newcommand{\los}{line of sight}

\newcommand{\am}{angular momentum}
\newcommand{\sam}{specific angular momentum}
\newcommand{\sams}{specific angular momenta}

\newcommand{\sysV}{systemic velocity}
\newcommand{\Eu}{$E_{\rm u}$}

\newcommand{\Msun}{$M_\odot$}

\newcommand{\rcb}{$r_{\rm CB}$}
\newcommand{\rCB}{\rcb}

\newcommand{\inv}{$^{-1}$}
\newcommand{\kmps}{km s\inv}
\newcommand{\fdegr}{.\!\!\degr}

\newcommand{\cmcubic}{cm$^{-3}$}

\newcommand{\Jypb}{Jy beam\inv}
\newcommand{\mJypb}{m\Jypb}
\newcommand{\ncrit}{$n_{\rm crit}$}

\newcommand{\distanceRem}{{\bff We employ 137 pc as the distance {\bf to} \iras\ \citep{Ortiz-Leon_Ophiuchus}.}}

\newcommand{\Vsys}{3.9 \kmps}
\newcommand{\PAenv}{50\degr}

\newcommand{\samNWpm}{${\displaystyle (9.0\pm0.3) \times 10^{-4}}$} 
\newcommand{\samSEpm}{${\displaystyle (8.8\pm0.2) \times 10^{-4}}$}

\newcommand{\samNWrangeIA}{$(8.7-14.3) \times 10^{-4}$}
\newcommand{\samSErangeIA}{$(8.6-14.1) \times 10^{-4}$}

\newcounter{tbnotecount}
\shorttitle{Rotation in the IRAS 16293-2422 A Outflow}
\shortauthors{Oya et al.}

\title{Rotating Motion of the Outflow of IRAS 16293--2422 A1 at its Origin Point near the Protostar}

\author{
Yoko Oya\altaffilmark{1, 2}, 
Yoshimasa Watanabe\altaffilmark{3}, 
Ana L\'opez-Sepulcre\altaffilmark{4, 5}, 
C\'ecilia Ceccarelli\altaffilmark{4}, 
Bertrand Lefloch\altaffilmark{4}, 
Cecile Favre\altaffilmark{4}, 
and Satoshi Yamamoto\altaffilmark{1, 2}
}

\email{oya@taurus.phys.s.u-tokyo.ac.jp}
\altaffiltext{1}{Department of Physics, The University of Tokyo, 7-3-1, Hongo, Bunkyo-ku, Tokyo 113-0033, Japan}
\altaffiltext{2}{Research Center for the Early Universe, The University of Tokyo, 7-3-1, Hongo, Bunkyo-ku, Tokyo 113-0033, Japan} 
\altaffiltext{3}{Materials Science and Engineering, College of Engineering, Shibaura Institute of Technology, 3-7-5 Toyosu, Koto-ku, Tokyo 135-8548, Japan}
\altaffiltext{4}{Universite Grenoble Alpes, CNRS, Institute de Plan\'etologie et d'Astrophysique de Grenoble (IPAG), 38000 Grenoble, Grance}
\altaffiltext{5}{Institut de Radioastronomie Millim\'etrique, 300 rue de la Piscine, Domaine Universitaire de Grenoble, 38406 Saint-Martin d'H\`eres, France}

\begin{abstract}
The Class 0 protostar IRAS 16293--2422 Source A is 
known to be a binary system (A1 and A2) or even a multiple system, 
which processes a complex outflow structure. 
We have observed this source in the C$^{34}$S, SO, and OCS lines at 3.1 mm 
with the Atacama Large Millimeter/submillimeter Array (ALMA). 
A substructure of this source is traced by 
our high angular-resolution observation (0\farcs12; 20 au) of the continuum emission. 
The northwest-southeast (NW-SE) outflow on a 2\arcsec\ scale 
is detected in the SO ($J_N$ = $2_2$--$1_1$) line. 
Based on the morphology of the SO distribution, 
this bipolar outflow structure seems to originate from the protostar A1 and its circumstellar disk, 
or the circummultiple structure of Source A. 
The rotation motion of the NW-SE outflow is detected in the SO and OCS emissions. 
We evaluate the specific angular momentum of the outflowing gas 
to be $(8.6 - 14.3) \times 10^{-4}$ km s$^{-1}$ pc. 
If the driving source of this outflow is the protostar A1 and its circumstellar disk, 
it can be a potential mechanism to extract the specific angular momentum of the disk structure. 
These results can be a hint for the outflow launching mechanism in this source. 
Furthermore, they provide us with an important clue to resolve 
the complicated structure of IRAS 16293--2422 Source A. 
\end{abstract}

\keywords{stars: protostars, stars: low-mass, ISM: jets and outflows, ISM: kinematics and dynamics, ISM: molecules, ISM: individual (IRAS 16293$-$2422)}

\begin{document}
\section{Introduction} \label{sec:intro}

In the last decade, 
the study of disk-formation around solar-type protostars 
has {\bff made} 
extensive progress 
both theoretically and observationally. 
In the disk-formation process around a newly-born protostar, 
outflows/jets and disks are mutually related via the angular momentum. 
However, 
their relation has not been elucidated in detail by observations. 
For instance, 
there is still difficulty to judge where outflows/jets are launched from; 
a central protostellar object, an inner-edge of a disk, or a disk surface. 
Moreover, 
the outflow launching process in multiple systems is expected to {\bff be complex.} 
Since a large fraction of stars are born as a member of a multiple or binary system 
\citep[e.g.][]{Chen2013, Duchene2013, Tobin2016}, 
investigating jet and outflow structures of binary/multiple systems 
is essential for the {\bff star-formation} processes ongoing there.

Outflows/jets from a binary/multiple system have extensively been studied 
by theoretical magnetohydrodynamics (MHD) simulations and observations. 
MHD simulations show that
an outflow of a binary system can be launched {\bff from} 
its circumbinary disk 
\citep[e.g.][]{Machida2009, Kuruwita2017}, 
and that twin jets can arise in a close binary system \citep{Saiki2020}. 
Specifically, in addition to the two jets from the two protostars in the latter case, 
a wide-angle and low-velocity outflow emanates 
from the structure composed by the circumbinary stream that encloses them. 
Observations of outflows/jets in binary/multiple systems are not always easy, 
and the identification of the outflow/jet launching points often requires high angular-resolution observations, 
particularly for close binary systems. 
For well-separated binaries, 
the outflow of each component is often identified, 
where the two outflows are sometimes parallel 
\citep[NGC 1333 IRAS 4A, BHR71;][]{Santangelo2015, Alves2017} 
and sometimes not 
\citep[L1448 IRS3, NGC 2264 CMM3;][]{Tobin2015, Watanabe2017}. 
In close binary systems, 
only one outflow is usually observed. 
For instance, 
\citet{Alves2017} reported a wide-angle outflow structure surrounding the binary system BHB 07--11, 
whose projected separation is 28 au \citep{Alves2019}. 
This seems to correspond to the low-velocity outflow seen in the recent MHD simulation by \citet{Saiki2020}. 
To the best of our knowledge, 
only one close binary system is known having misaligned twin molecular outflows \citep{Hara2020}: 
VLA 1623A, whose two binary objects are separated by 34 au. 
Considering the importance of outflows/jets in the formation and evolution of disk structures, 
high {\bff angular-resolution} observations of various binary/multiple sources are of fundamental importance. 

\iras\ is a Class 0 protostellar source 
located in Ophiuchus, 
{\bff whose distance from the Solar system is 
reported to be $(137-147)$ pc by \citet{Ortiz-Leon_Ophiuchus} 
and $141^{+30}_{-21}$ pc by \citet{Dzib2018}.
In this study, 
we employ 137 pc as the distance {\bf to} \iras\ \citep{Ortiz-Leon_Ophiuchus} 
for the consistency with the previous analysis \citep[e.g.][]{Oya_16293, Oya_16293Cyc4}.} 
This source contains two sources (Source A and Source B), 
which are separated by $\sim5\arcsec$ ($\sim700$ au) on the \pos\ 
\citep[e.g.][]{Wootten1989, Mundy1992, Looney2000, Chandler2005, Oya_16293, Jorgensen_PILS}. 
Both \iras\ Source A and Source B are known to be prototypical hot corino sources, 
which are rich in complex organic molecules (COMs), 
{\bff such as \MN\ (methanol), \MF\ (methyl formate), \DME\ (dimethyl ether), and \PPN\ (propionitrile) 
\citep[e.g.][]{vanDishoeck1995, Cazaux2003, Schoier_hotcore, Bottinelli2004, Kuan2004, Ceccarelli2004, Huang2005, Chandler2005, Caux2011, Jorgensen2011, Jorgensen_sugar}.} 
Their chemical compositions have extensively been investigated 
with {\bf the} {\bff Submillimeter Array \citep[SMA;][]{Jorgensen2011}
and {\bf the} Atacama Large Millimeter/submillimeter Array (ALMA) 
\citep[e.g. PILS; Protostellar Interferometric Line Survey;][]{Jorgensen_PILS, Lykke2017} 
as well as with the IRAM 30 m telescope {\bf and the James Clerk Maxwell Telescope (JCMT)} \citep{Caux2011}. 
} 

\iras\ Source A itself is a binary source, or even a multiple source, 
including at least two protostars A1 and A2 
\citep[][, hereafter \citetalias{Oya_16293Cyc4}]{Wootten1989, Huang2005, Chandler2005, Pech2010, HernandezGomez2019_contVLA, Maureira2020, Oya_16293Cyc4}. 
The projected spatial separation between the protostar A1 and A2 is as small as 
0\farcs36 \citep[$\sim50$ au; e.g.][; \citetalias{Oya_16293Cyc4}]{HernandezGomez2019_contVLA, Maureira2020}. 
Moreover, 
Source A is reported to have substructures other than A1 and A2 
\citep[][; \citetalias{Oya_16293Cyc4}]{Pech2010, HernandezGomez2019_contVLA}. 
\citet{Loinard2013}, \citet{Pech2010}, and \citet{HernandezGomez2019_contVLA} 
observed the cm continuum emission with Very Large Array (VLA), 
and reported a bipolar ejection from the protostar A2. 

In this study, 
we investigate the outflow feature observed toward Source A at a high {\bff angular-resolution}, 
corresponding to about 20 au. 
In Section \ref{sec:intro_16293outflow}, 
previous studies of the outflows in \iras\ are summarized, 
and the principal aim of this paper is presented. 
Section \ref{sec:obs} describes the observation details, 
and Section \ref{sec:results} the results and discussions, 
where 
we particularly focus on the rotation motion of the outflow (Sections \ref{sec:results_kin_outflow} and \ref{sec:disc_sam}). 
{\bff Finally, in Section \ref{sec:summary}, 
we summarize our results.}

\section{Outflows of \iras} \label{sec:intro_16293outflow}

The outflow structure of \iras\ is quite complicated 
\citep[see e.g. the review by][]{vanderWiel2019}. 
\citet{Mizuno1990} and \citet{Stark2004} reported a quadruple outflow structure on a 6\arcmin\ ($\sim0.24$ pc) scale 
extending along the east-west {\bf (E-W)} and the northeast-southwest {\bf (NE-SW)} directions. 
\citet{Walker1988} and \citet{Hirano2001} observed a similar structure over 2\arcmin\ ($\sim0.08$ pc). 
\citet{Castets2001} reported multiple shocks caused by these outflows. 
On a smaller scale ($\sim10^3$ au scale), 
outflow structures from the individual components (Source A and Source B) have been observed. 
A nearly pole-on outflow structure of Source B was suggested by \citet{Loinard2013} and \citet{Oya_16293B}. 
Meanwhile, 
Source A is known to have two bipolar outflow structures on a $10^3$ au scale. 
\citet{Yeh2008} detected the {\bf E-W} outflow in the CO (\coYeha; \coYehb) lines, 
and \citet{Loinard2013} reported that the CO (\coLoinard) line traces the central part of this outflow within 1\arcsec\ around Source A. 
\citet{vanderWiel2019} delineated the northwest-southeast (NW-SE) outflow 
in the SiO ($J$ = 8--7) and H$^{13}$CN ($J$ = 4--3) lines. 
Both the E-W outflow and {\bff the} NW-SE outflow 
were detected by 
\citet{Rao2009} 
{\bff in} 
the CO (\coRao) line, the SiO ($J$ = 8--7) line, {\bff and} the H$^{13}$CO$^+$ ($J$ = 4--3) line, 
{\bff by} \citet{Kristensen2013} {\bff in} the CO (\coKristensen) line, 
and {\bff by} \citet{Girart2014} {\bff in} the CO (\coGirart) line {\bff and} the SiO ($J$ = 8--7) line. 

The E-W outflow on a $10^3$ au scale is blue-shifted and red-shifted on the eastern and western sides of Source A, respectively 
\citep{Yeh2008, Rao2009, Kristensen2013, Girart2014}. 
This feature is the same as that seen on the arcminute-scale outflow structure 
\citep{Walker1988, Mizuno1990, Stark2004}. 
The NW-SE outflow is blue-shifted and red-shifted on the {\bf NW} and {\bf SE} sides of Source A, respectively, 
at the distance of $\sim5\arcsec$ ($\sim700$ au) from Source A \citep{Rao2009, Girart2014, vanderWiel2019}. 
Interestingly, 
it shows the opposite feature at the distance of $\sim1\arcsec$ ($\sim140$ au) from Source A \citep{Kristensen2013}.

The E-W outflow has been reported to originate from the protostar A2 
based on the proper motion study of the bipolar ejecta from A2 in the cm continuum emission 
\citep{Loinard2013, Pech2010, HernandezGomez2019_contVLA}. 
On the contrary, 
the driving source of the NW-SE outflow is still unclear.  
In this study, 
we focus on the NW-SE outflow structure of \iras\ Source A, 
and present new high angular-resolution ($\sim20$ au) observations 
resolving the substructure in Source A with ALMA. 
We investigate the kinematic structure of this outflow near its launching point 
by using the \CS, SO, and OCS lines.

\section{Observation} \label{sec:obs}

The ALMA observation of \iras\ was carried out on November 16th and 28th, 2017 
during its Cycle 5 operation (\#2017.1.01013.S). 
The 3.1 mm continuum and 
the rotational spectral lines of \CS\ (\cs), SO (\so), and OCS (\ocsa) 
were observed with the Band 3 receiver. 
The observed spectral lines are summarized in Table \ref{tb:lines}.

In this observation, 
the field center 
was set at 
$(\alpha_{\rm ICRS}, \delta_{\rm ICRS}) = (16^{\rm h}32^{\rm m}22\fs79, -24\degr28\arcmin34\farcs30)$, 
which is the intermediate position of \iras\ Source A and Source B. 
Forty-three antennas were used during the observations, 
and their baseline lengths ranged from 92 to 8282 m. 
The size of the field of view 
was from 61\arcsec\ to 71\arcsec\ 
and the maximum recoverable scale was from 2\farcs3 to 2\farcs9 
according to the quality assurance report,  
both of which depend on the frequency.  
The on-source integration time was 113 minutes in total. 
Seven spectral windows shown in Table \ref{tb:tuning} were observed. 
The bandpass and flux calibrations were performed with J1427--4206. 
The phase calibration was carried out with J1633--2557 every 12 minutes. 
The accuracy of the flux density calibration for the target source image is typically expected to be 1\%\ \citep{ALMA_TH-C5}. 

The continuum and line images were obtained with the CLEAN algorithm 
by using {\tt CASA} 
{\bff \citep[Common Astronomy Software Applications package;][]{McMullin_CASA}.} 
We employed the Briggs weighting with a robustness parameter of 0.5, 
{\bff unless otherwise noted.} 
The 3.1 mm continuum image was prepared by averaging line-free channels 
with a cumulative frequency-range of 65.43 MHz. 
The line images were obtained after subtracting the continuum component directly from the visibility data. 
The line images were resampled to make the channel width to be 0.2 \kmps. 
A primary beam correction was applied to the continuum and line images. 
Self-calibration was carried out for the phase and amplitude by using the continuum data and was applied to the line data. 
The synthesized beam size and the root-mean-square (rms) noise level 
for the continuum image or each molecular line image 
are summarized in Table \ref{tb:tuning}. 

\section{Results and Discussions} \label{sec:results}

\subsection{Distributions} \label{sec:results_distribution}

Figure \ref{fig:mom0}(a) shows the 3.1 mm continuum image. 
The continuum emission shows an elliptical distribution with a major-axis diameter of $\sim$ 1\farcs5 ($\sim$ 210 au) 
extending along the {\bf NE-SW} direction. 
It can be regarded to trace the circummultiple \desys\ of \iras\ Source A, 
whose mid-plane is reported to extend along the position angle (P.A.) of \PAenv\ \citepalias{Oya_16293Cyc4}. 
Two intensity peaks (A1 and A2) are clearly seen. 
Their coordinates are summarized in Table \ref{tb:cont}, 
which well correspond to those found in previous observations 
\citep[e.g.][; \citetalias{Oya_16293Cyc4}]{Wootten1989, Chandler2005, Pech2010, HernandezGomez2019_contVLA, Maureira2020}. 
In addition, 
three faint intensity peaks are found by using the {\tt 2D Fit Tool} of {\tt CASA viewer} (Table \ref{tb:cont}) 
with an intensity excess {\bff less than $2 \sigma$ (Figure \ref{fig:cont-spatialprofile}). 
Even with the uniform weighting, 
the intensity excess is less than $3 \sigma$,} 
as shown in the spatial profiles along the lines passing through the continuum peak positions. 
Two {\bff of} 
them are close to the peaks A3 and A4 detected in the 1.3 mm continuum image \citepalias{Oya_16293Cyc4}. 
The bipolar ejecta from the protostar A2 found in the cm observation 
\citep[A2$\alpha$, A2$\beta$;][]{HernandezGomez2019_contVLA} 
are not detected in the 3.1 mm continuum image. 

Figures \ref{fig:mom0}(b), (c), and (d) show 
the integrated intensity maps of the \CS\ (\cs), SO (\so), and OCS (\ocsa) lines. 
The distribution of the \CS\ emission is similar to that of the 3.1 mm continuum emission (Figure \ref{fig:mom0}b). 
On the contrary, 
the SO emission 
protrudes from the continuum distribution toward the {\bf NW} and {\bf SE} directions (Figure \ref{fig:mom0}c). 
This elongation is perpendicular to the mid-plane of the \desys\ (P.A. \PAenv). 
{\bff On this scale,} 
the SO emission morphologically seems to trace 
the bipolar outflow blowing along the {\bf NW-SE} direction 
previously reported by \citet{Rao2009}, \citet{Kristensen2013}, \citet{Girart2014}, and \citet{vanderWiel2019}. 
The OCS emission is slightly more extended along the \desys\ than the 3.1 mm continuum image 
and the \CS\ emission (Figure \ref{fig:mom0}d). 
It would partly trace the NW-SE outflow structure in addition to the extended envelope gas.

The difference in the morphological size of the emitting region of the observed molecular lines 
is probably due to 
spatial differentiation of molecular abundances, 
as well as
the excitation conditions of the used lines. 
The critical density (\ncrit) for the excitation is evaluated {\bff to be 
$\sim (4-5) \times 10^5$ \cmcubic\ for the \CS\ (\cs) line,} 
$\sim 2 \times 10^5$ \cmcubic\ for the SO (\so) line, 
and $\sim 2 \times 10^4$ \cmcubic\ for the OCS (\ocsa) line 
{\bff by using their Einstein A coefficients and state-to-state collisional rate coefficients 
\citep{Yamamoto2017}.} 
The collisional rate coefficients 
are originally reported by 
\citet{Lique2006a, Lique2006b} and \citet{Green1978} 
for the SO, \CS, and OCS lines, respectively. 
We use the values which are 
{\bff calculated based on} 
the original data and summarized 
in Leiden Atomic and Molecular Database \citep[LAMDA;][]{Schoier_LAMDA}. 
The Einstein A coefficients are also taken from LAMDA, 
 {\bff except for the one of the \CS\ (\cs) line which is calculated from its line parameters (Table \ref{tb:lines}). 
Here, we assume that the collisional rate coefficients 
of \CS\ can be approximated by those of C$^{32}$S.} 

In addition to the NW-SE outflow, 
\iras\ Source A is known to have an outflow blowing along the {\bf E-W} direction, 
according to the CO (\coGirart) observations by \citet{Girart2014}. 
Such a structure is not evident in our SO and OCS observations {\bf (Figures \ref{fig:mom0}c, d)} 
{\bf and the reason for the non-detection is not clear.} 
{\bf This might be due to} 
the difference of the molecular abundances 
which {\bf are} 
highly 
{\bff dependent on the physico-chemical {\bf conditions} and} {\bf the} age of {\bf the} shocks 
\citep{Wakelam2004}. 
In addition, 
the excitation effect may contribute to the non-detection of the {\bf E-W} outflow. 
The SO (\so) line has a higher critical density than 
the CO (\coGirart) line \citep[\ncrit\ $= (2 \times 10^3 - 2 \times 10^4)$ \cmcubic;][]{Yang2010}, 
while the OCS (\ocsa) line has a similar critical density to the CO (\coGirart) line. 
A combination of these two effects as well as a resolving-out effect 
in the interferometric observation would be a cause of the non-detection. 

\subsection{Kinematic Structures of the Molecular Lines} \label{sec:results_kinematics} 

Figure \ref{fig:mom1} shows the velocity (moment 1) maps of the \CS\ (\cs), SO (\so), and OCS (\ocsa) lines. 
The \CS\ line clearly traces a velocity gradient along the {\bf NE-SW} direction (Figure \ref{fig:mom1}a). 
This is regarded as the rotating motion in the \desys\ of \iras\ Source A reported previously 
\citep[e.g.][; \citetalias{Oya_16293Cyc4}]{Pineda_ALMA, Favre_SMA, Oya_16293, Maureira2020}. 
It seems to trace the circummultiple structure surrounding both the protostars A1 and A2 \citepalias{Oya_16293Cyc4}.

\citetalias{Oya_16293Cyc4} have recently reported that 
\iras\ Source A has a circummultiple structure surrounding both the protostars A1 and A2 
and a circumstellar disk associated with the protostar A1. 
Figure \ref{fig:scheme_sam} shows the schematic illustration for Source A. 
According to their result, 
the innermost radius of the circummultiple structure is 50 au, 
and the circumstellar disk resides inside it. 
It is found that 
the \CO\ (\co) emission mainly tracing the circummultiple structure shows a hole in its distribution around the protostar A1. 
In contrast, 
the distribution of the \CS\ emission shown in Figure \ref{fig:mom0}(b) extends over 200 au in diameter, 
without {\bf the} depression near its central position. 
Thus, the \CS\ emission seems to trace both the circummultiple and circumstellar structures.

\subsection{Rotating Motion of the Outflow} \label{sec:results_kin_outflow}

As shown in Section \ref{sec:results_distribution}, 
the SO emission is extended along the NW-SE outflow 
beyond the circummultiple structure traced by the continuum and \CS\ emission. 
{\bff 
Morphologically speaking, 
the SO emission seems to trace the outflow, or its cavity wall instead, 
although it may also trace the surface of a flared envelope gas. 
In fact, 
the SO emission has been reported to trace a disk wind in another protostellar system \citep[HH 212;][]{Tabone2017, Lee2021}.
}

{\bff The SO emission shows} 
a velocity gradient along the NE-SW direction (Figure \ref{fig:mom1}b), 
which is almost perpendicular to the NW-SE outflow axis. 
A similar gradient across the NW-SE outflow is also seen in the OCS line (Figure \ref{fig:mom1}c). 
These features are schematically illustrated in Figure \ref{fig:mom1}(d).

In order to investigate the kinematic structure of the SO (\so) line,  
we show the integrated intensity maps for every 4 \kmps\ in Figure \ref{fig:mom0_4kmps}. 
Figure \ref{fig:mom0_4kmps}(a) compares the map integrated from 0 to 4 \kmps\ and that from 4 to 8 \kmps, 
which are blue-shifted and red-shifted, respectively, 
with respect to the \sysV\ (\Vsys) 
of the circummultiple structure \citep[][, \citetalias{Oya_16293Cyc4}]{Bottinelli2004}. 
Figure \ref{fig:mom0_4kmps}(b) shows the maps with higher velocity-shift components 
($-4-0$ \kmps\ and $8-12$ \kmps). 
Although the protostar A2 {\bff is located} within the outflow structure 
{\bff as shown in Figure \ref{fig:mom0_4kmps}(a),} 
it is slightly offset from the outflow axis. 
Thus, the outflow structure seems to be associated with the protostar A1 rather than A2 morphologically. 
Alternatively, 
the outflow may be a disk wind from the circummultiple structure (Figure \ref{fig:scheme_sam}). 
{\bff The latter} 
is examined based on the \sam\ of the outflow in Section \ref{sec:disc_sam_compare}.

Figures \ref{fig:mom0_4kmps}(b) and (c) show 
the high velocity-shift ($-4-0$ and $8-12$ \kmps) components near the protostars A1 and A2. 
These components would mostly trace the rotating motion of the circummultiple/circumstellar structures. 
On the other hand, 
the high-velocity blue-shifted {\bff component ($-4-0$ \kmps)} 
on the SE side of the protostar A1 in Figure \ref{fig:mom0_4kmps}(b) 
is naturally interpreted as a part of the outflow structure. 
Since this component is near the outflow axis, 
its velocity-shift would be due to the outflowing motion.  
The NW outflow lobe shows both the red- and blue-shifted high-velocity components. 

Figures \ref{fig:mom0_4kmps}(a) and (c) clearly show 
that 
the {\bf NE} edges of the bipolar outflow lobes are blue-shifted 
while their {\bf SW} edges are red-shifted. 
This situation is schematically illustrated in Figure \ref{fig:mom1}(d). 
Such a velocity structure strongly suggests a velocity gradient across the outflow lobes. 
The gradient can indeed be seen in the velocity {\bff (moment 1)} maps 
for the SO and OCS lines {\bff (Figures \ref{fig:mom1}b, c).} 
{\bff We can also confirm a clear velocity gradient 
in the position-velocity (PV) diagrams of the SO emission (Figure \ref{fig:PV}). 
The position axes of the diagrams are taken across the outflow axis. 
A velocity gradient along the NE-SW direction is clearly seen 
in both the NW and SE outflow lobes; 
the NE parts of the two lobes are blue-shifted, 
while the SW parts are red-shifted. 
This is consistent with what we see in Figure \ref{fig:mom0_4kmps}. 
}

Such features in outflows have been reported, for instance, 
for the Source I in Orion Kleinmann-Low \citep{Hirota2017}, L483 \citep{Oya_483outflow}, NGC 1333 IRAS 4C \citep{Zhang2018}, 
{\bff and HH 212 \citep{Tabone2020, Lee2021}.} 
They are interpreted as the rotating motion of the outflows, 
where the outflow blows nearly perpendicular to the \los. 
It is most likely that 
the gradient seen in the outflow part of the SO and OCS emission (Figures \ref{fig:mom1}b, c) in the NW-SE outflow is 
ascribed to its rotating motion. 

\subsection{Direction of the Outflow} \label{sec:disc_direction} 

Since the NW-SE outflow blows along the direction nearly perpendicular to the \los, 
the velocity structure near the protostar shows a complex feature due to the contribution 
from both of the outflowing motion and the rotating motion of the outflow. 
To investigate the direction of the NW-SE outflow lobes, 
we {\bff consider} the spectra of SO at a few positions (Figure \ref{fig:spectra}). 
We confirm that the SO line in our observation is free from contaminations by other molecular lines according to molecular databases 
\citep[JPL and CDMS;][]{Pickett_JPL, Muller_CDMS}, 
although this hot corino is rich in COM lines. 
Figure \ref{fig:spectra}(b) shows the spectra on the outflow axis in the NW and SE outflow lobes. 
In Figure \ref{fig:mom0_4kmps}(b), 
the NW lobe shows both red- and blue-shifted components, 
while the SE lobe shows only a blue-shifted component. 
As well, both red- and blue-shifted components are clearly confirmed for the NW lobe in Figure \ref{fig:spectra}(b). 
For the SE lobe, 
we {\bff detect} a red-shifted component weakly but certainly in addition to a clear blue-shifted component, 
although they overlap each other. 
The velocity centroids of these four components on the outflow axis 
are evaluated by the Gaussian fitting (Table \ref{tb:gaussfit_spectra}). 
The velocity centroids are more blue-shifted in the SE lobe than in the NW lobe. 

The more blue-shifted component on the SE side of Source A 
is consistent with the report by \citet{Kristensen2013}, 
where this component was detected in the CO (\coKristensen) line within 1\arcsec\ ($\sim140$ au) around Source A. 
As described in Section \ref{sec:intro_16293outflow}, 
the NW-SE outflow shows the opposite direction on a larger scale 
\citep[$\sim5\arcsec$, $\sim 700$ au;][]{Rao2009, Girart2014, vanderWiel2019}. 
A part of the previous observations show 
that the NW-SE outflow lobes have both blue-shifted and red-shifted components 
as shown in our observation; 
the SiO observation by \citet{Rao2009} (their Figure 6), 
the H$_2$CO, H$_2^{13}$CO, SiO, and H$^{13}$CN lines by \citet{vanderWiel2019} (their Figure 3). 
\citet{Loinard2013} and \citet{Girart2014} 
also detected the red- and blue-shifted components on the NW and SE sides of Source A, respectively. 
This trend 
is consistent with our observation and the CO observation by \citet{Kristensen2013}. 

Combining previous {\bff observations with ours,} 
the NW-SE outflow seems to blow almost {\bff in} parallel to the \pos\ 
in the vicinity of Source A on a $<$500 au scale. 
In this case, 
both the red- and blue-shifted components come from the back and front sides of the expanding outflow structure. 
The outflow axis may slightly be tilted from the \pos, 
where the NW and SE lobes blow away from and toward us, respectively. 

In addition, 
the outflow may have precession 
and its direction can vary according to the distance from Source A. 
Since the \desys\ of Source A has a complex structure, 
temporal variation of the outflow direction may occur. 
These situations make it further complicated to discuss the direction of the outflow axis. 
Nevertheless, 
the conclusion that the NW-SE outflow blows almost parallel to the \pos\ is robust 
and is consistent with the nearly edge-on circummultiple/circumstellar structures of this source \citepalias{Oya_16293Cyc4}.

{\bff One may think that the SO (\so) emission comes from the root of the E-W outflow rather than the NW-SE outflow 
as suggested by \citet{Loinard2013} and \citet{Girart2014}. 
The E-W outflow is thought to cause the bipolar ejecta (A2$\alpha$, A2$\beta$) 
detected with VLA observations \citep{HernandezGomez2019_contVLA}. 
The positions of the ejecta are shown in Figure \ref{fig:mom0_4kmps}(c). 
A2$\alpha$ is out of our SO emission, although A2$\beta$ is within it; 
the ejecta are located on the eastern and western sides of Source A, 
while the SO emission extends along the NW-SE direction from Source A. 
Considering their positional relation, 
our SO emission does not likely trace the root of the E-W outflow on this scale, 
but the NW-SE outflow. 
}


\subsection{Specific Angular Momentum} \label{sec:disc_sam}

\subsubsection{Launching Radius of the NW-SE Outflow Lobes} \label{sec:disc_launch-radius}

Figures \ref{fig:spectra}(c) and (d) {\bff are} the spectra on the edges of the outflow lobes. 
We {\bff show} the spectra toward the four positions depicted in Figure \ref{fig:spectra}(a). 
These spectra are fitted by Gaussian profiles, 
and the fitting results including their velocity centroids 
are summarized in Table \ref{tb:gaussfit_spectra}. 
{\bff The velocity gradient inferred from the obtained velocities at Positions C and D is shown by a dashed line in Figure \ref{fig:PV}(a), 
while that at Positions E and F is shown 
in Figure \ref{fig:PV}(b). 
The dashed lines well trace the overall trend in the {\bf position-velocity} diagrams.}

{\bff The} outflowing 
{\bff motion should equally contribute to the velocity centroids at} 
the NE and SW edges in {\bff each} outflow lobe, 
{\bff so that its contribution}
can be eliminated approximately by taking the difference of the velocity centroids between the {\bff two} edges. 
{\bff As well, a possible infall motion can be eliminated.} 
{\bff Thus, the rotation velocity of an outflow lobe is obtained as: 
\begin{equation}
v_{\rm rot} = \frac{\left| v_{\rm NE} - v_{\rm SW} \right|}{2 \sin i}, \label{eq:vrot}
\end{equation}
where $v_{\rm NE}$ and $v_{\rm SW}$ denote the velocity centroids of the spectra on the NE and SW edges, respectively, 
and $i$ the \ia\ of the outflow (\incRemOutflow). 
Using the derived velocity difference (Table \ref{tb:gaussfit_spectra}; Figure \ref{fig:spectra}), 
$v_{\rm rot}$ is evaluated to be $2.25\pm0.07$ \kmps\ and $2.21\pm0.06$ \kmps\ for the NW and SE lobes, respectively. 
}

{\bff 
The launching radius of the outflowing gas is often derived from its rotating motion \citep[e.g.][]{Zhang2018}. 
\citet{Anderson2003} gave the equation to derive a launching radius of a magneto-centrifugal wind ($\varpi_0$)
(their equation (5)): 
{\bff 
\begin{equation}
\varpi_0 \sim 0.7 \left( \frac{\varpi_\infty}{10\ {\rm au}} \right)^{2/3} \left( \frac{v_{\phi, \infty}}{10\ {\rm km\ s}^{-1}} \right)^{2/3} 
\left( \frac{v_{p, \infty}}{100\ {\rm km\ s}^{-1}} \right)^{-4/3} \left( \frac{M}{1\ M_\odot} \right)^{1/3} \ {\rm au}, 
\label{eq:launch-radius}
\end{equation}
where 
$\varpi_\infty$, $v_{\phi, \infty}$, and $v_{p, \infty}$ denote 
the distance from the outflow axis, the rotation velocity, and the poloidal velocity 
at a position far from the launching point, respectively, 
while $M$ denotes the protostellar mass.} 
{\bff Since 
the poloidal velocity of the outflow of \iras\ Source A is highly uncertain due to 
the outflow direction almost parallel to the \pos, 
we here assume a large variety of} 
a poloidal velocity of 3, 10, and 30 \kmps\ as its reasonable range. 
{\bff Then,} 
the launching radius of the NW outflow lobe of \iras\ Source A is calculated to be 
$83\pm2$, $16.7\pm0.4$, and $3.87\pm0.08$ au, respectively. 
Here, we employed the rotation velocity of the outflow derived above 
and the protostellar mass of 0.4 \Msun\ for the protostar A1 \citepalias{Oya_16293Cyc4}. 
Employing the central mass of 1.0 \Msun\ for Source A \citepalias{Oya_16293Cyc4} instead, 
the launching radius is calculated to be $133\pm2$, $22.7\pm0.5$, and $5.3\pm0.1$ au. 
Unfortunately, 
{\bff due to the high uncertainties of the poloidal velocity of the outflow and the protostellar mass,} 
the launching radius is not well constrained by this method. 
{\bff The estimation obtained here only gives the practical upper limit of 133 au. 
Its improvement is left for future study.} 
Therefore, 
we just discuss whether the outflow can contribute to 
the extraction of the \am\ from the \desys\ in the following sections. 
}

\subsubsection{Specific Angular Momentum of the NW-SE Outflow Lobes} \label{sec:disc_sam_outflow}

The \sam\ ($j$) of the gas is related to the rotation velocity 
and the distance ($d$) between the NE and SW edges of an outflow lobe, 
i.e. the diameter of the outflow lobe as: 
\begin{equation}
j = v_{\rm rot} \times \frac{d}{2}. \label{eq:j_outflow}
\end{equation}
{\bff Using the rotation velocity ($v_{\rm rot}$) obtained in Section \ref{sec:disc_launch-radius},}  
we evaluate the \sam\ of the gas in the NW and SE outflow lobes rotating around the outflow axis to be 
\samNWpm\ \kmps\ pc and 
\samSEpm\ \kmps\ pc, respectively, 
where the outflow lobes are assumed to be parallel to the \pos\ ($i$ = 90\degr).

We here note a caveat on 
the uncertainty of the \ia\ of the outflow. 
If the outflow lobes are inclined from the \pos\ ($i <$ 90\degr), 
the above values of the \sam\ should be divided by $\sin i$ as described in the equation (\ref{eq:vrot}). 
Thus, they should be regarded as underestimations. 
According to \citetalias{Oya_16293Cyc4}, 
the reasonable range of the \ia\ is 
from 70\degr\ to 90\degr\ for the circummultiple structure 
and from 40\degr\ to 70\degr\  for the circumstellar disk. 
Employing these ranges of the \ia, 
the \sam\ of the outflow lobes are modified, as summarized in Table \ref{tb:sam}. 

{\bff 
We also perform a similar analysis for the OCS (\ocsa) line. 
The details for the results are described in Appendix \ref{sec:app_OCS}. 
The velocity gradients across the outflow lobes found in the SO emission are confirmed in the OCS emission, 
and thus the derived \sams\ are almost consistent with those derived from the SO emission 
within a factor of 2 (See Appendix \ref{sec:app_OCS} for more detailed comparison). 
}

\subsubsection{Specific Angular Momentum of the Circummultiple Structure of Source A and the Circumstellar disk of the Protostar A1} \label{sec:disc_sam_desys}

\citetalias{Oya_16293Cyc4} have recently reported 
the kinematic structure of the circummultiple system of Source A and that of the circumstellar disk around the protostar A1 
at a spatial resolution of $\sim$ 0\farcs1 (14 au). 
The \sam\ of these two structures can be evaluated by using the physical parameters 
(the protostellar mass, the \ia, and the radius of the \cb) 
obtained in their model analysis. 

According to \citetalias{Oya_16293Cyc4}, 
the circummultiple structure of Source A is reproduced 
by an envelope model with a constant \sam. 
The \sam\ ($j$) is obtained by using 
the protostellar mass ($M$) and the radius of the \cb\ (\rCB) as: 
\begin{equation}
	j = \sqrt{2 G M r_{\rm CB}}, \nonumber
\end{equation}
where $G$ denotes the gravitational constant \citep{Oya_15398}. 
The results are summarized in Table \ref{tb:sam}. 

Meanwhile, 
the circumstellar disk of the protostar A1 is reported to be reproduced 
by a Keplerian disk model. 
If the gas has a Keplerian rotation with a radius of $r$ from the central protostar with the mass of $M$, 
its \sam\ is obtained as: 
\begin{equation}
	j = \sqrt{r G M}. \nonumber
\end{equation}
The radius of the circumstellar disk is a bit uncertain. 
It is 30 au based on the \TFA\ observation, 
while its upper limit is 50 au based on the inner most radius of the circummultiple structure surrounding the circumstellar disk. 
Hence, we evaluate the \sam\ for the three cases (the radius of 10, 30, and 50 au), 
considering the uncertainty of the launching point of the outflow. 
Again, the results are summarized in Table \ref{tb:sam}. 

It should be noted that the physical parameters of the models employed in the above calculations 
are highly correlated with each other as demonstrated 
by \citetalias{Oya_16293Cyc4} (See Tables A1--4 in their paper). 
In Table \ref{tb:sam}, 
the \sam\ values are corrected for the effect of the \ia, 
and the uncertainty of the other physical parameters are taken into account.

\subsubsection{Comparison between the Outflow and the Circummultiple/circumstellar Structures} \label{sec:disc_sam_compare}

The evaluated \sam\ of the NW-SE outflow is 
larger than that of the circumstellar disk of the protostar A1 (Table \ref{tb:sam}). 
{\bff In Appendix \ref{sec:app_OCS}, 
we confirm that this conclusion does not change critically, 
even if we employ the OCS line instead of the SO line. 
The conclusion does not change either for the reference positions of $\pm$1\arcsec\ instead of $\pm$0\farcs6. 
Hence, the result is quite robust.} 
If the driving source of this outflow structure is the protostar A1 and its associated disk, 
the outflow likely plays an important role in extracting the \sam\ of the disk structure, 
which allows the gas to accrete onto the protostar. 

To investigate the \am\ transportation more quantitatively, 
we need to consider the transportation of the gas mass in addition to the \sam. 
\citet{Maureira2020} have recently reported the gas mass of 
$(0.001 - 0.003)$ \Msun\ and $(0.03-0.1)$ \Msun, 
for the substructure around the protostar A1 and the extended structure surrounding both of A1 and A2, respectively. 
On the other hand, it is difficult to evaluate the gas mass of the outflow from our observation result. 
Therefore, we present only a little thought to the balance of the \am\ between the outflow and the \desys\ with rough assumptions. 
If the gas mass of the outflow were similar to that of the circumstellar disk, 
the NW-SE outflow could extract $(12 - 61)$ \%\ of \sam\ of the infalling gas onto the disk 
(See Appendix \ref{sec:app_angmom} for details). 
Thus, the NW-SE outflow can be one of the mechanisms for the \am\ loss of the circumstellar disk. 

Alternatively, 
the outflow may be a disk wind from the circummultiple structure of Source A (Section \ref{sec:results_kin_outflow}). 
For instance, 
\citet{Bjerkeli2016c} reported the CO observation tracing the outflow of the Class I low-mass protostellar source TMC-1A; 
the wide-angle outflow structure was interpreted as an extended disk wind. 
In our observation, the SO and OCS lines may trace a similar structure 
to the disk wind of TMC-1A (Figures \ref{fig:mom0}, \ref{fig:mom1}). 
In Table \ref{tb:sam}, 
the \sam\ of the outflow seems slightly smaller than that of the circummultiple structure, 
and thus, the outflow may not play an effective role to extract the \sam\ of the infalling gas. 
However, the above estimations of the \sam\ suffer from uncertainties of the physical parameters, 
and it would be too hasty to conclude with the current observational results (See Appendix \ref{sec:app_angmom}).


{\bff 
It should be noted that 
the observed SO distribution may represent the outflow cavity wall rather than the outflowing gas itself, 
as mentioned in Section \ref{sec:results_kin_outflow}. 
If this is the case, 
the observed rotating velocity can be smaller than the actual velocity of the outflowing gas, 
which leads underestimation of its \sam. 
Then, 
the outflow may have a \sam\ larger than both the circummultiple structure and the circumstellar disk; 
the outflow can play a role in extracting the angular momentum from the accreting gas, 
wherever in the \desys\ the outflow is launched. 
}

As shown in Table \ref{tb:sam}, 
the \sam\ is different 
between the circummultiple structure and the circumstellar disk inside it in this source. 
If the NW-SE outflow does not play a significant role in the extraction of the \am\ from the cicummultiple structure, 
another mechanism needs to be considered to account for the observed difference; 
for instance, 
a possible transportation of the angular momentum between a binary and its circumbinary disk 
\citep[e.g.][]{Miranda2017, Moody2019, Tiede2020, Heath2020}. 
This is left for future study.

\section{Summary} \label{sec:summary} 

We have observed the \CS, SO, and OCS line {\bf emission} as well as the 3.1 mm continuum emission 
with ALMA 
toward the Class 0 {\bff protostar} \iras\ Source A at a resolution from 0\farcs12 to 0\farcs2 (from 20 to 30 au). 
We have investigated the kinematic structure of the {\bf NW-SE} outflow 
by analyzing the velocity structure of the SO line. 
The major findings are as follows.

\begin{itemize}
\item[(1)] 
The substructure of \iras\ Source A is delineated in the 3.1 mm continuum emission. 
The protostars A1 and A2 are clearly detected. 
The continuum emission also shows an extended distribution along the {\bf NE-SW} direction, 
corresponding to the nearly edge-on \desys\ of this source known previously.

\item[(2)] 
The SO (\so) emission traces 
a bipolar outflow structure extending along the NW-SE direction from Source A 
on a 2\arcsec\ ($\sim$ 300 au) scale. 
The \CS\ (\cs) line traces the rotating disk structure of Source A. 
It is likely the combination of the circummultiple structure of Source A and 
the circumstellar disk of the protostar A1. 
The OCS (\ocsa) line seems to trace a part of the NW-SE outflow 
as well as the circummultiple structure and the circumstellar disk. 

\item[(3)]
The NW-SE outflow blows almost in parallel to the \pos, 
although its NW and SE lobes would be slightly red- and blue-shifted, respectively. 

\item[(4)]
The NW-SE outflow does not originate from the protostar A2 
based on its morphology, 
but 
likely from the protostar A1 and its circumstellar disk, or the circummultiple structure. 

\item[(5)]
The NW-SE outflow shows a rotating motion in our SO observation. 
Its \sam\ is evaluated to be \samNWrangeIA\ \kmps\ pc and \samSErangeIA\ \kmps\ pc
for the NW and SE lobes, respectively, 
considering the range of \ia\ from 40\degr\ to 90\degr\ (\incRemOutflow). 
These values are larger than that of the circumstellar disk of the protostar A1. 
Thus, this outflow can play a role in the extraction of the \sam\ from the disk structure, 
if its driving source is the protostar A1 and its circumstellar disk. 
Although 
the NW-SE outflow does not seem to extract the \sam\ from the circummultiple structure of Source A 
significantly based on the current observational results, 
we need more {\bff  accurate} 
observations for a definitive conclusion. 

\end{itemize}

\acknowledgements 

The authors acknowledge Dr. Aya Higuchi for her invaluable help in the data analysis. 
The authors are grateful to Dr. Yichen Zhang for his invaluable discussions. 
The authors also thank to Dr. Nami Sakai for her helpful comments and discussions in preparation of the proposal for the observation. 
This study uses the ALMA data set ADS/JAO.ALMA
\#2017.1.01013.S. 
ALMA is a partnership of the European Southern Observatory, 
the National Science Foundation (USA), 
the National Institutes of Natural Sciences (Japan), 
the National Research Council (Canada), 
and the NSC and ASIAA (Taiwan), 
in cooperation with the Republic of Chile. 
The Joint ALMA Observatory is operated by the ESO, the AUI/NRAO, and the NAOJ. 
The authors are grateful to the ALMA staff for their excellent support. 
This study is supported by a Grant-in-Aid from the Ministry of Education, 
Culture, Sports, Science, and Technologies of Japan 
(grant Nos. 18H05222, 19H05069, 19K14753, {\bff 21K13954).} 
This project has received funding from the European Research Council (ERC) 
under the European Union's Horizon 2020 research and innovation programme, 
for the Project “The Dawn of Organic Chemistry” (DOC), grant agreement No. 741002.

\appendix

\section{Specific Angular Momentum Analysis Using the OCS (\ocsa) Line} \label{sec:app_OCS} 

{\bff 
In Section \ref{sec:disc_sam}, 
we analyze the velocity structure of the SO (\so) line. 
In this Section, we perform a similar analysis for the OCS (\ocsa) line. 

Figure \ref{fig:spectra_OCS} shows the spectra of the OCS line. 
Since the OCS emission is more extended than the SO emission, 
we obtain the spectra at four more positions in addition to the six positions employed for the SO analysis. 
The additional positions (Positions C\arcmin, D\arcmin, E\arcmin, F\arcmin) are taken at the distance of 1\farcs0 (137 au) from Position A or B. 
The spectra at the ten positions are fitted by Gaussian profiles, 
and the fitting results are summarized in Table \ref{tb:gaussfit_spectra_OCS}. 
By using their velocity centroids, 
we calculate the \sam\ of the gas (Table \ref{tb:sam_OCS}). 
Since the spectra at Positions C and E show a double peak profile, 
they are fitted by using two Gaussian profiles. 
Then, the average of the two velocity centroids weighted by the peak intensity is calculated for the velocity at Positions C and E, 
and is used for the angular momentum calculation. 
As described in Section \ref{sec:disc_launch-radius}, 
the contamination of the outflowing motion of the gas is expected to be canceled out 
in the calculation of the \sam\ of the gas (equations (\ref{eq:vrot}) and (\ref{eq:j_outflow})), 
while the contribution of the \ire\ may remain. 

For the NW outflow lobe, 
the \sam\ calculated by using the OCS line is slightly larger than that calculated by using the SO line 
by 5 up to 6$\sigma$ at the distance of 0\farcs6 (80 au) from the outflow axis (Tables \ref{tb:sam}, \ref{tb:sam_OCS}). 
Although the spectrum at Position E\arcmin\ has a low signal-to-noise ratio, 
the \sam\ obtained for the SE outflow lobe at the distance of 1\farcs0 (137 au) from the outflow axis 
agrees with those obtained in the SO analysis within their uncertainties. 

Meanwhile, 
the \sams\ obtained at the distance of 1\farcs0 in the NW lobe and at that of 0\farcs6 in the SE lobe 
tend to be lower than those obtained in the SO analysis. 
Since the OCS emission likely traces the \ire\ as well as the NW-SE outflow as described in Section \ref{sec:results_distribution}, 
the \sams\ obtained based on the OCS emission are likely subject to the contamination of the envelope gas. 
The contribution of the envelope gas would cause a difference 
between the \sams\ obtained in the OCS and SO analyses. 
In fact, 
the relatively small \sams\ obtained in the OCS analysis just correspond to the intermediate value 
between the those obtained in the SO analysis and the \sam\ of the disk/envelope structures. 

Despite the uncertainties mentioned above, 
the \sams\ obtained in the NW-SE outflow lobes tend to be 
larger than that of the circumstellar disk around the protostar A1, 
and are likely smaller than that of the circummultiple structure of Source A. 
A few exceptions happen 
when we employ the \sam\ of the circumstellar disk around the protostar A1 
calculated at its upper limit radius (50 au; Table \ref{tb:sam}). 
However, 
it does not change the main conclusion that 
the NW-SE outflow lobe can contribute to the \am\ extraction from the circumstellar disk 
considering 
that its launching position should be within the outermost edge of the disk 
(Sections \ref{sec:disc_sam_compare}, \ref{sec:summary}), 
}

\section{Balance of the Angular Momentum} \label{sec:app_angmom}

When a gas particle falls toward the protostar, 
it cannot fall inward of its periastron due to the centrifugal force. 
Thus, infalling gas needs to lose its \sam\ for the protostellar evolution. 
Outflow launching is thought to be one of the candidate mechanism for the \am\ loss of infalling gas. 

The balance of the \am\ between an outflow and an infalling gas can be formulated as reported by \citet{Oya_483outflow}. 
We consider the case that an infalling gas splits into two gas clumps with a gas mass of $m_1$ and $m_2$. 
Then, the balance of the \am\ among these three gas components is described as: 
\begin{equation}
	(m_1 + m_2) j_0 = m_1 j_1 + m_2 j_2, \label{eq:am_balance}
\end{equation}
where $j_0$, $j_1$, and $j_2$ denote {\bff the \sams\ of} the infalling gas and the two gas clumps. 

Here, we suppose that the gas clump with the mass of $m_1$ and the \sam\ of $j_1$ to be the outflowing gas. 
We obtain the ratio of {\bff the \sams\ between} the other two gas clumps as: 
\begin{equation}
	\frac{j_2}{j_0} = \frac{m_1 + m_2}{k m_1 + m_2}, \label{eq:am_ratio} 
\end{equation}
{\bff where $j_1 = k j_2$.} 
If $k$ is larger than 1, 
the \sam\ of an infalling gas decreases from $j_0$ to $j_2$ by the outflow launching. 

In our observation (Section \ref{sec:disc_sam}), 
we evaluate $j_{\rm 1}$ to be \samNWrangeIA\ \kmps\ pc and \samSErangeIA\ \kmps\ pc 
for the {\bf NW} and {\bf SE} outflow lobes, respectively, 
considering the uncertainty of the \ia. 
We also {\bff obtained} the \sam\ of the circummultiple structure of Source A 
and the circumstellar disk of the protostar A1 
as shown in Table \ref{tb:sam}.  

If the outflow launches from the circumstellar disk, 
$k$ is evaluated to be from 1.3 to 4.1. 
\citet{Maureira2020} have recently reported the gas mass of 
$(0.001 - 0.003)$ \Msun\ for the substructure around the protostar A1, 
and this can be employed as $m_2$. 
If the gas mass of outflow ($m_1$) were the same as $m_2$, 
${\displaystyle \frac{j_2}{j_0}}$ would be obtained to be from 0.39 to 0.88  
by using the equation (\ref{eq:am_ratio}). 
In other words, 
the infalling gas loses from 12 to 61 \%\ of its \sam\ ($j_0$) by outflow launching, 
and falls near the protostar A1 to form its circumstellar disk. 
The infalling gas loses more \sam\ for a larger gas mass of outflow, 
and the upper limit {\bff to} of 
the \sam\ loss is obtained to be 76 \%\ {\bff ${\displaystyle \left( 1 - \frac{1}{k} \right)}$.} 

The \sam\ of the circummultiple structure seems to be larger than that of the NW-SE outflow. 
If this is the case, 
$k$ is smaller than 1, and thus ${\displaystyle \frac{j_2}{j_0}}$ is larger than 1. 
In other words, 
the NW-SE outflow does not seem to extract the \sam\ from the circummultiple structure significantly. 
Nevertheless, 
the evaluations of the \sam\ in our observations would have a large uncertainties. 
As well, 
it should be noted that 
$m_0$ and $m_2$ depends on where the outflow is launched. 
This is {\bff also} highly uncertain with our observations. 
Therefore, the conclusion 
{\bff for the \am\ extraction from the circummultiple structure} 
is left for future study.

\newcommand{\iffigure}{\iftrue}

\newcommand{\dirfig}{} 

\begin{table}
\begin{center}
	\caption{Molecular Lines\tablenotemark{a} 
			\label{tb:lines}}
	\vspace*{10pt}
	\begin{tabular}{lcccc}
	\hline 
	Molecule & Transition & Rest Frequency & S$\mu^2$  & \Eu \vspace*{-10pt} \\ 
	& & (GHz) & ($D^2$) & (K) \\ \hline \hline 
	3 mm Continuum & & $97.2331-98.1696$ & & \\ 
	\CS & \cs & 96.41294950  & 7.6 & 6.2 \\ 
	SO & \so & 86.09395000 & 3.5 & 19 \\ 
	OCS & \ocsa & 85.13910320 & 3.6 & 16 \\ 
	\hline 
	\end{tabular}
	\tablenotetext{a}{Taken from CDMS \citep{Muller_CDMS}.}
\end{center}
\end{table}

\begin{landscape}
\begin{table}
	\begin{center}
	\caption{Settings of the Spectral Windows in the Observation
			\label{tb:tuning}}
	\begin{tabular}{ccccccc}
	\hline
	SPW ID & Frequency Range & Resolution & Molecular Lines & Beam Size & RMS\tablenotemark{a} \vspace*{-10pt} \\ 
	& (GHz) & (kHz) & & & (\mJypb) \\ \hline \hline 
	0\tablenotemark{b} & 84.5675--84.5089 & 30.517 & \vspace*{-10pt} \\ 
	\\ 
	1 & 85.1795--85.1209 & 30.517 & OCS (\ocsa) & 0\farcs250 $\times$ 0\farcs199 & 3 \vspace*{-10pt} \\ 
	& & & & (P.A. $-71\fdegr925$) & \\ 
	2 & 86.1284--86.0699 & 61.035 & SO (\so) & 0\farcs245 $\times$ 0\farcs173 & 2 \vspace*{-10pt} \\ 
	& & & & (P.A. $-54\fdegr152$) & \\ 
	3\tablenotemark{b} & 86.8754--86.8169 & 61.035 & \vspace*{-10pt} \\ 
	\\ 
	4 & 96.3827--96.4413 & 30.517 & \CS\ (\cs) & 0\farcs211 $\times$ 0\farcs167 & 3 \vspace*{-10pt} \\ 
	& & & & (P.A. $-49\fdegr137$) & \\  
	5\tablenotemark{b} & 97.271--97.3296 & 61.035 & \vspace*{-10pt} \\ 
	\\ 
	6 & 97.2331--98.1696 & 976.555 & Continuum & 0\farcs136 $\times$ 0\farcs112 & 0.09 \vspace*{-10pt} \\ 
	& & & & (P.A. $64\fdegr080$) & \\ 
	\hline 
	\end{tabular}
	\tablenotetext{a}{Root-mean-square noise level in each image. The line images have a velocity channel width of 0.2 \kmps.} 
	\tablenotetext{b}{Not used in this study.} 
	\end{center}
\end{table}
\end{landscape}

\begin{table}
	\begin{center}
	\caption{Intensity Peaks in the 3.1 mm Continuum Map\tablenotemark{a}
			\label{tb:cont}}
	\begin{tabular}{lccc}
	\hline 
	Peak & Peak Intensity (mJy/beam) & RA (ICRS) & Dec (ICRS) \\ \hline \hline 
	A1 & 5.98 & 16$^{\rm h}$32$^{\rm m}$22\fs879 & $-$24\arcdeg28\arcmin36\farcs691 \\ 
	A2 & 4.27 & 16$^{\rm h}$32$^{\rm m}$22\fs851 & $-$24\arcdeg28\arcmin36\farcs660 \\ 
	A3?\tablenotemark{b} & 2.05 & 16$^{\rm h}$32$^{\rm m}$22\fs887 & $-$24\arcdeg28\arcmin36\farcs493 \\ 
	A4?\tablenotemark{b} & 1.93 & 16$^{\rm h}$32$^{\rm m}$22\fs847 & $-$24\arcdeg28\arcmin36\farcs802 \\ 
	A5?\tablenotemark{b} & 1.37 & 16$^{\rm h}$32$^{\rm m}$22\fs889 & $-$24\arcdeg28\arcmin36\farcs266 \\  
	\hline 
	\end{tabular}
	\tablenotetext{a}{Obtained from the continuum image {\bff for the Briggs weighting of 0.5} 
				by using the {\tt 2D Fit Tool} of {\tt casa viewer}.} 
	\tablenotetext{b}{Tentative detection with an intensity excess less than 2$\sigma$.} 
	\end{center}
\end{table}

\begin{table}
\begin{center}
	\caption{Results of the Gaussian Fitting for the Spectra of the SO (\so) Line\tablenotemark{a}
			\label{tb:gaussfit_spectra}}
	\vspace*{10pt}
	\begin{tabular}{lcccc}
	\hline \vspace*{-5pt} 
	Position & Distance from & Peak Intensity & Velocity Centroid & FWHM\tablenotemark{b}  \\ \vspace*{3pt} 
	& the Outflow Axis & (\mJypb) & (\kmps) & (\kmps)  \\ \hline \hline 
	\multicolumn{5}{c}{Northwestern Outflow Lobe} \\ \hline 
	Center (Position A) & 0\arcsec & $15.6 \pm 0.7$ & $2.5 \pm 0.1$ & $5.3 \pm 0.3$ \\ 
	& & $9.2 \pm 1.0$ & $8.7 \pm 0.1$ & $2.2 \pm 0.3$ \\ 
	NE Edge (Position C) & 0\farcs6 (80 au) & $11.7 \pm 0.7$ & $0.7 \pm 0.1$ & $4.2 \pm 0.3$ \\ 
	SW Edge (Position D) & 0\farcs6 (80 au) & $7.4 \pm 0.8$ & $5.2 \pm 0.1$ & $2.2 \pm 0.3$ \\ 
	\hline 
	\multicolumn{5}{c}{Southeastern Outflow Lobe} \\ \hline 
	Center (Position B) & 0\arcsec & $13.8 \pm 0.5$ & $1.1 \pm 0.1$ & $5.0 \pm 0.3$ \\ 
	& & $5.2 \pm 0.9$ & $5.5 \pm 0.2$ & $2.0 \pm 0.4$ \\ 
	NE Edge (Position E) & 0\farcs6 (80 au) & $8.4 \pm 0.6$ & $1.1 \pm 0.1$ & $3.9 \pm 0.3$ \\ 
	SW Edge (Position F) & 0\farcs6 (80 au) & $13.1 \pm 0.9$ & $5.52 \pm 0.07$ & $2.2 \pm 0.2$ \\ 
	\hline
	\end{tabular}

	\tablenotetext{a}{To derive the line parameters, each spectrum is taken 
					at each position
					depicted in Figure \ref{fig:spectra}(a).
					\distanceRem  
					} 
	\tablenotetext{b}{Full width at half maximum of the Gaussian profile.} 
\end{center}
\end{table}

\begin{landscape}
\begin{table}
\begin{center}
	\vspace*{-60pt}
	\caption{Specific Angular Momenta of the Outflow Lobes and the Circummultiple/circumstellar Structures\tablenotemark{a} 
			\label{tb:sam}}
	\vspace*{-10pt}
	\begin{tabular}{lccccccc}
	\hline 
	Structure & Distance from & \multicolumn{6}{c}{Inclination Angle\tablenotemark{b}} \\ 
	& the Outflow Axis & 40\degr & 50\degr & 60\degr & 70\degr & 80\degr & 90\degr \\ \hline \hline
	NW Outflow Lobe & 80 au & $13.9\pm0.4$ & $11.7\pm0.4$ & $10.4\pm0.3$ & $9.5\pm0.3$ & $9.1\pm0.3$ & $9.0\pm0.3$ \\ 
	SE Outflow Lobe & 80 au & $13.7\pm0.4$ & $11.5\pm0.3$ & $10.2\pm0.3$ & $9.4\pm0.3$ & $8.9\pm0.2$ & $8.8\pm0.2$ \\ 
	\begin{tabular}{c} \hspace*{-10pt} \vspace*{-5pt} Circummultiple Structure \\ of Source A\tablenotemark{c} \end{tabular}
	& $50-300$ au & $-$ & $-$ & $-$ & $14.4$ & $12.9-15.8$ & $11.2-15.8$ \\ 
	\multirow{2}{*}{\begin{tabular}{c} \hspace*{-10pt} \vspace*{-5pt} Circumstellar Disk \\ of Protostar A1\tablenotemark{d} \end{tabular}} 
	& 10 au & $3.5-4.1$ & $2.9-3.5$ & $2.9-3.2$ & $2.9-3.2$ & $-$ & $-$ \\ 
	& 30 au & $6.1-7.1$ & $5.0-6.1$ & $5.0-5.6$ & $5.0-5.6$ & $-$ & $-$ \\ 
	& 50 au & $7.9-9.1$ & $6.5-7.9$ & $6.5-7.2$ & $6.5-7.2$ & $-$ & $-$ \\ 
	\hline 
	\end{tabular}

\vspace*{-30pt}

	\tablenotetext{a}{The \sams\ are in the unit of $10^{-4}$ \kmps\ pc. 
					{\bff The values in the unit of \kmps\ au can be obtained by 
					multiplying those in \kmps\ pc by $2.06 \times 10^5$.} 
					The ranges of the \sams\ for the outflow lobes are 
					based on the uncertainties calculated by the propagation from the fitting errors of the velocity centroids (Table \ref{tb:gaussfit_spectra}). 
					\distanceRem 
					} 
	\tablenotetext{b}{\incRemOutflow.}
	\tablenotetext{c}{The cicummultiple structure is reproduced by the \ire\ model, 
					where the \sam\ of the gas conserves everywhere. 
					The \sam\ is calculated by employing the protostellar mass and the radius of the \cb\ 
					reported by \citetalias{Oya_16293Cyc4}. 
					The lower and upper limits of the \ia\ {\bff are} 70\degr\ and 90\degr, respectively. 
					The best-fit parameters in \citetalias{Oya_16293Cyc4} are the protostellar mass of 1.0 \Msun, the radius of the \cb\ of 50 au, and \ia\ of 80\degr, 
					where the \sam\ of the gas is calculated to be $14.4 \times 10^{-4}$ \kmps\ pc.} 
	\tablenotetext{d}{The circumstellar disk of the protostar A1 is reproduced by the Keplerian model, 
					where the \sam\ of the gas increases as the distance from the protostar.  
					The \sam\ is calculated for three radii 
					by employing the protostellar mass reported by \citetalias{Oya_16293Cyc4}. 
					The lower and upper limits of the \ia\ {\bff are} 40\degr\ and 70\degr, respectively. 
					The best-fit parameters in \citetalias{Oya_16293Cyc4} are the protostellar mass of 0.4 \Msun\ and the \ia\ of 60\degr, 
					where the \sam\ of the gas is calculated to be 
					$2.9 \times 10^{-4}$, $5.0 \times 10^{-4}$, and $6.5 \times 10^{-4}$ \kmps\ pc 
					at the radius of 10, 30, and 50 au, respectively. 
					} 
\end{center}
\end{table}
\end{landscape}

\begin{figure}
\begin{center}
	\iffigure
	\epsscale{1.0}
	\plotone{\dirfig 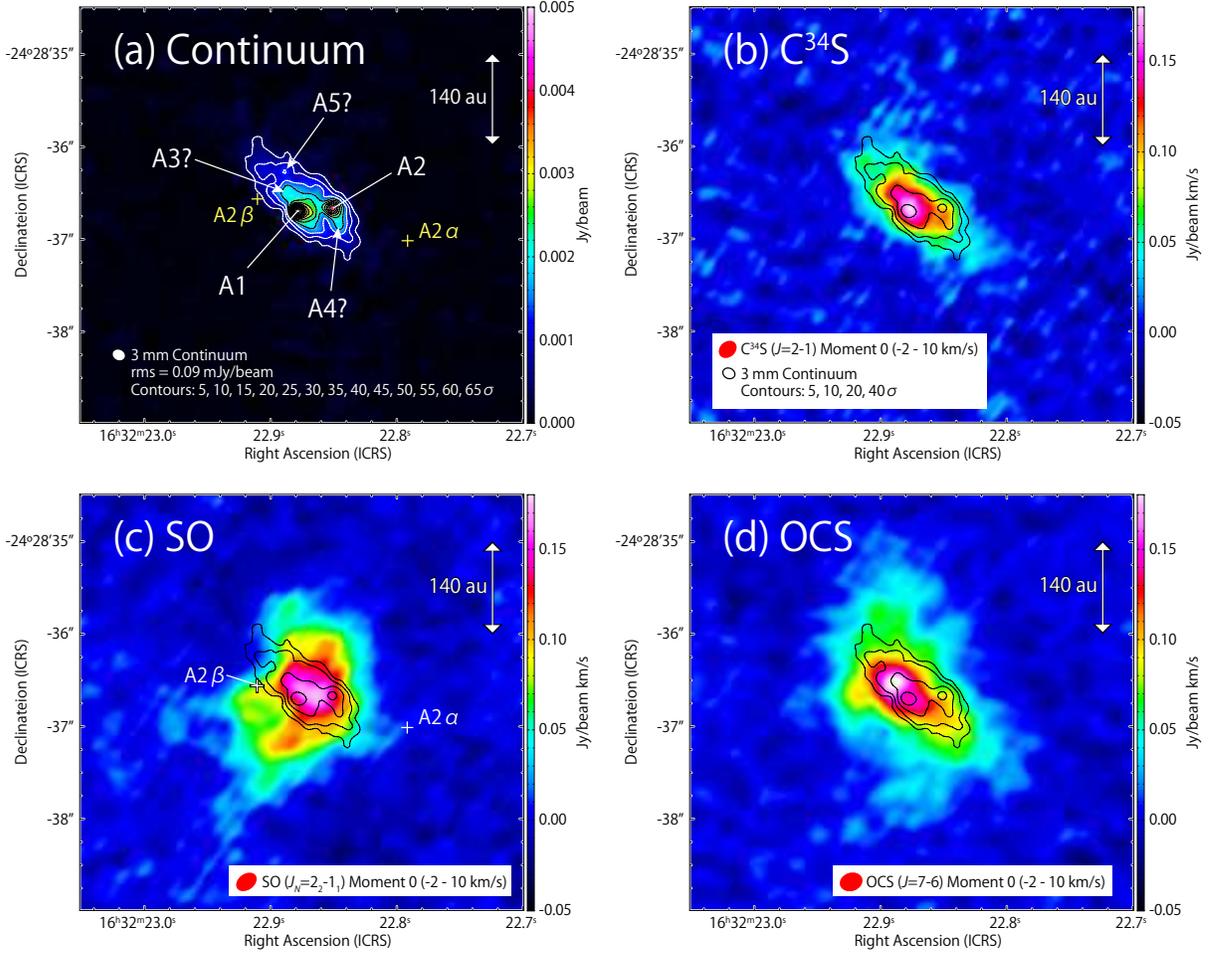} 
	\fi
	\caption{The 3.1 mm continuum image (a) and the integrated intensity maps of 
			the \CS\ (\cs; b), SO (\so; c), and OCS (\ocsa; d) lines. 
			Contours represent the continuum image. 
			Contour levels are every 5$\sigma$ in panel (a), 
			while they are 5, 10, 20, and 40$\sigma$ in the other panels. 
			The rms noise level is 0.09 \mJypb. 
			In panels (b), (c), and (d), 
			the velocity range for the integration is from $-2$ to 10 \kmps, 
			which corresponds to the velocity shift from $-5.9$ to $+6.1$ \kmps\ 
			for the \sysV\ of the circummultiple structure (\Vsys). 
			Two crosses in panel (a) show the expected positions of 
			the two ejecta (A2$\alpha$ and A2$\beta$) on our 1st observation date (2017.88) 			
			estimated by extrapolating their proper motions from their positions on 2014.15 
			reported by \citet{HernandezGomez2019_contVLA}. 
			\label{fig:mom0}}
\end{center}
\end{figure}

\begin{figure}
\begin{center}
	\iffigure
	\epsscale{1.0}
	\plotone{\dirfig 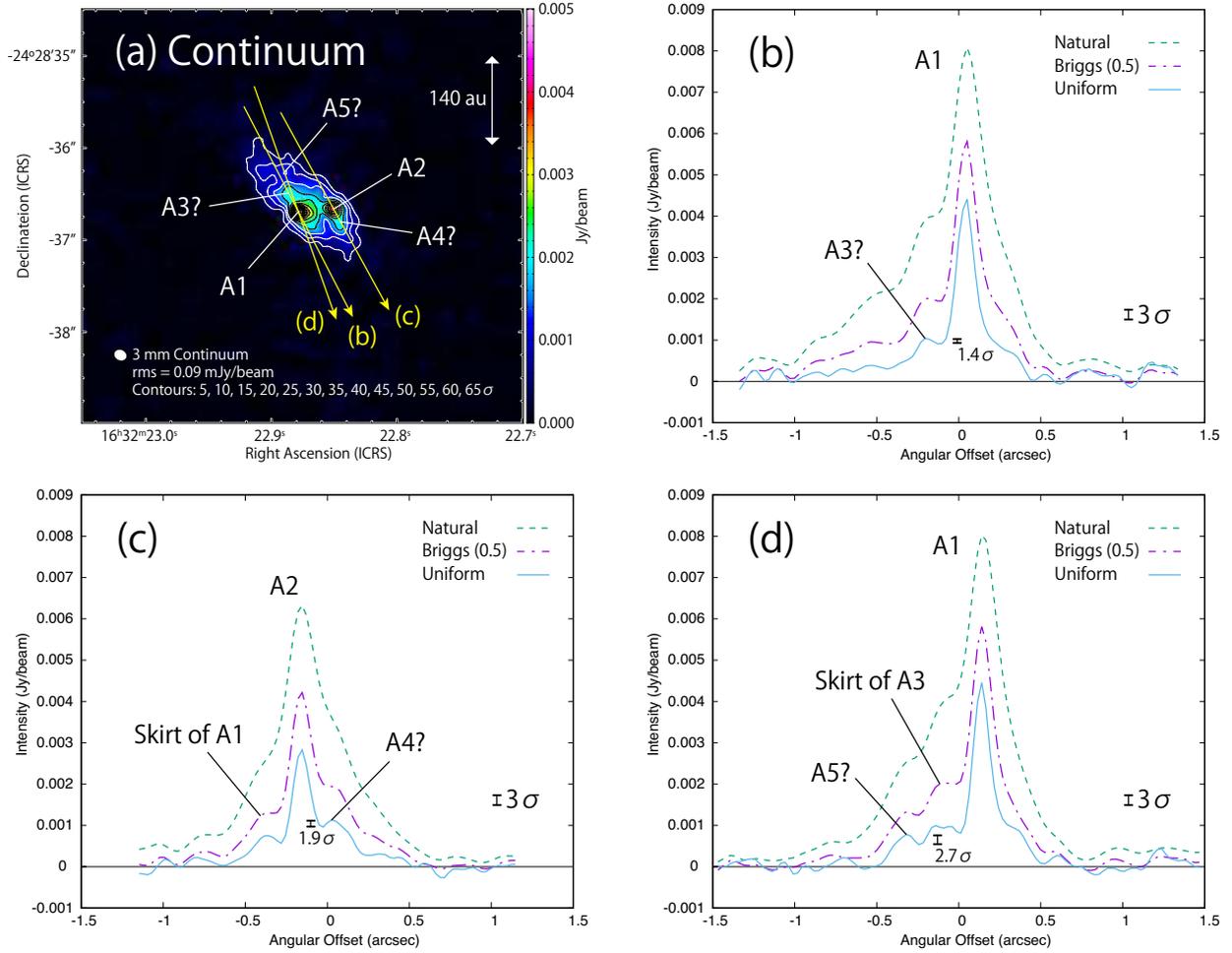}
	\fi
	\caption{The 3.1 mm continuum image (a) and its spatial profiles (b--d). 
			Panel (a) is the same as panel (a) in Figure \ref{fig:mom0}. 
			Spatial profiles are prepared along the arrows shown in panel (a). 
			The arrows pass through the positions of the intensity peak candidates. 
			{\bff The spatial profiles are prepared for the continuum images 
			with the robustness parameter of $+2$ (natural), 0.5, and $-2$ (uniform). 
			The beam size is 
			0\farcs197 $\times$ 0\farcs162, 	
			0\farcs136 $\times$ 0\farcs112, 	
			and 0\farcs114 $\times$ 0\farcs070, 	
			with the robustness parameter of $+2$, 0.5, and $-2$, respectively. 
			The three intensity peaks (A3, A4, and A5) 
			are tentatively detected by less than $3 \sigma$ in the image with the uniform weighting. 
			A scale of the 3 $\sigma$ noise level of the image with the uniform weighting is depicted in the bottom left corner in each panel, 
			where the rms noise level is 0.9 \mJypb. 
			} 
			\label{fig:cont-spatialprofile}}
\end{center}
\end{figure}

\begin{figure}
\begin{center}
	\iffigure
	\epsscale{1.0}
	\plotone{\dirfig 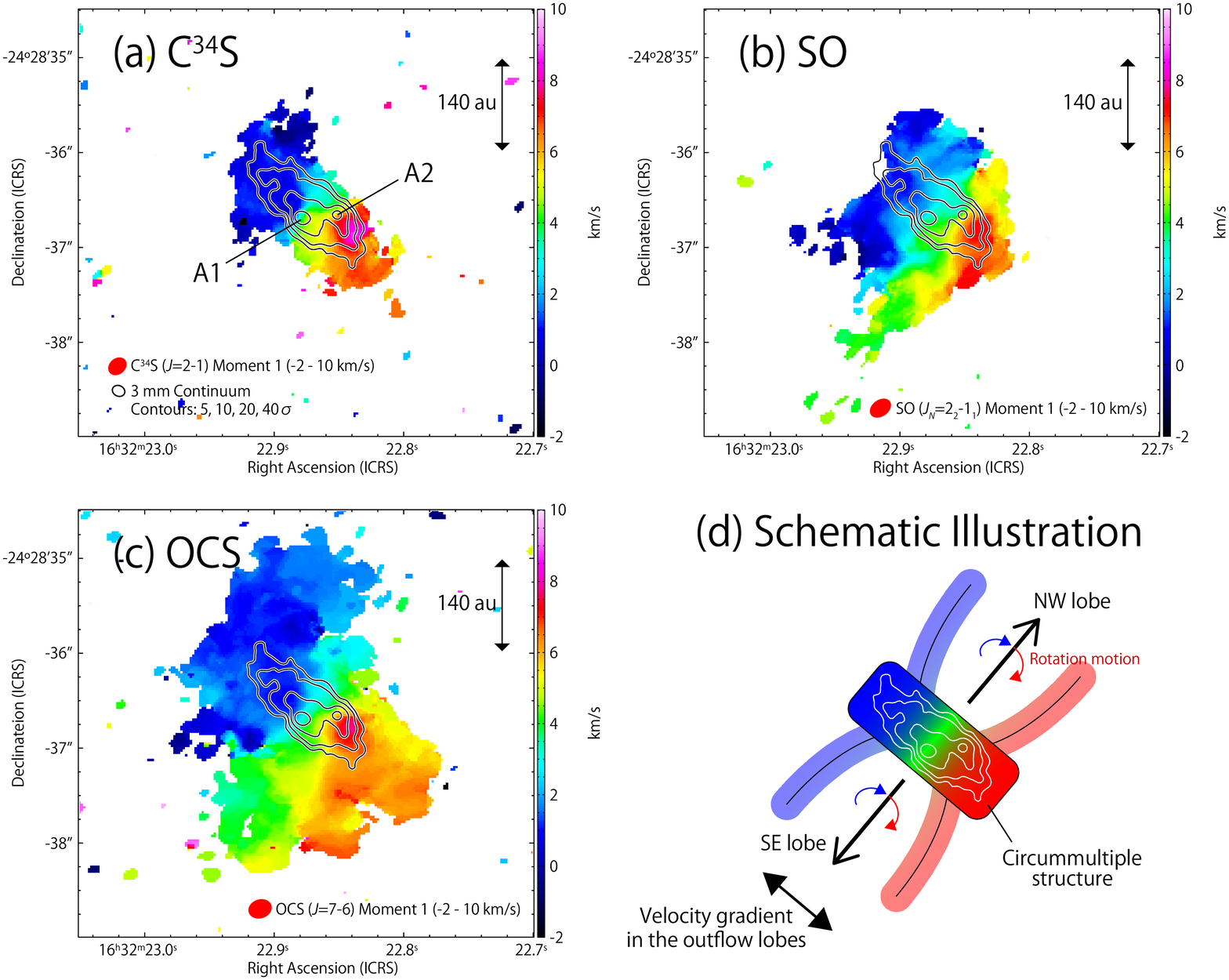} 
	\fi
	\caption{Velocity maps (moment 1 maps) of 
			the \CS\ (\cs; a), SO (\so; b), and OCS (\ocsa; c) lines. 
			Contours represent the 3.1 mm continuum image, 
			where the contour levels are as the same as those in panels (b), (c), and (d) in Figure \ref{fig:mom0}. 
			The velocity range for the integration is from $-2$ to 10 \kmps, 
			which corresponds to the velocity shift from $-5.9$ to $+6.1$ \kmps\ 
			for the \sysV\ of \Vsys. 
			The color maps are prepared by using the data points 
			with the intensity higher than 10 \mJypb\ in the {\bf datacube;} 
			10 \mJypb\ corresponds to 5$\sigma$ for the SO line  
			and to $\sim$3$\sigma$ for the \CS\ and OCS lines. 
			(d) Schematic illustration for the velocity distribution. 
			\label{fig:mom1}}
\end{center}
\end{figure}

\begin{figure}
\begin{center}
	\iffigure
	\epsscale{0.9}
	\plotone{\dirfig 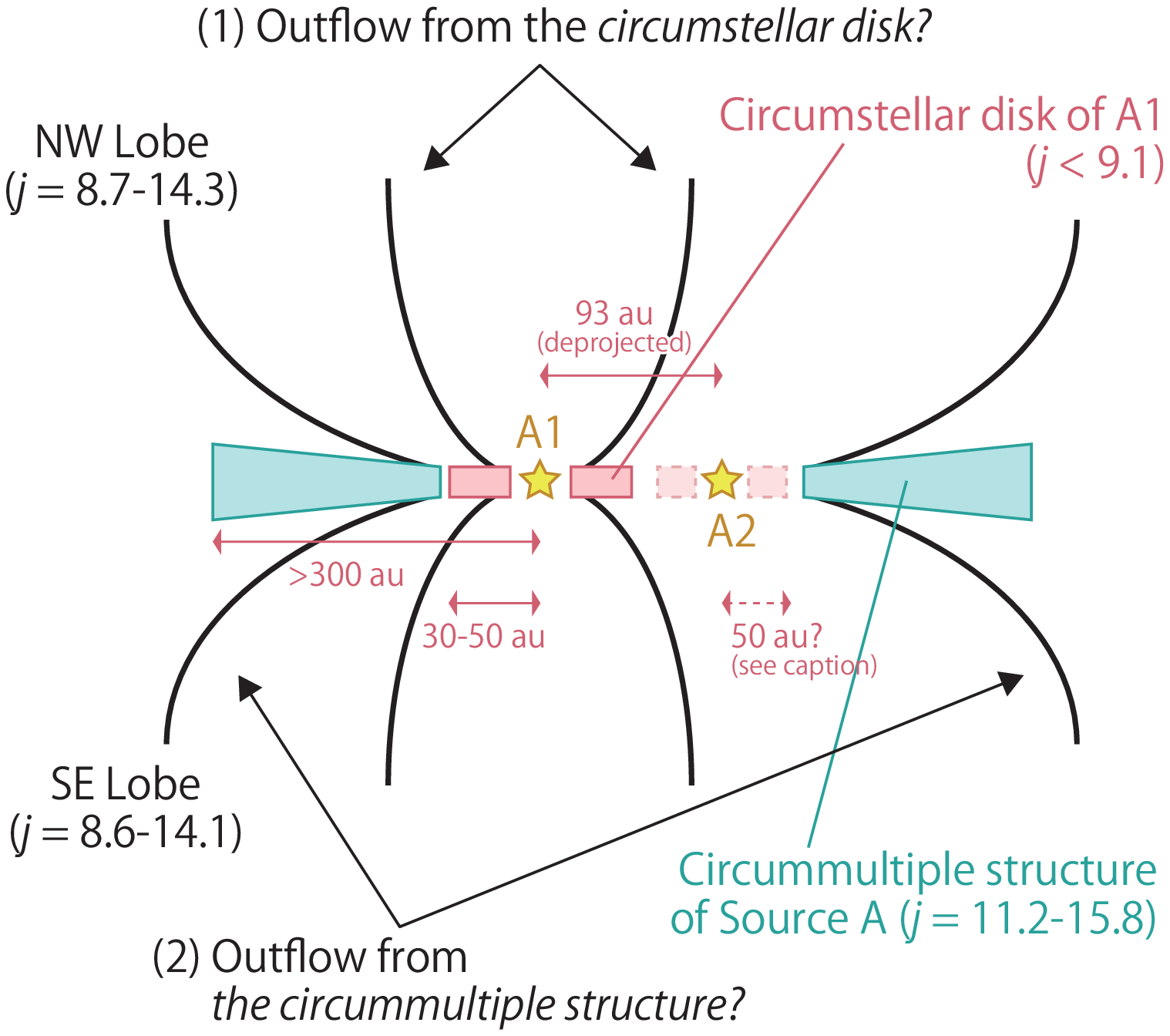}
	\fi
\end{center}
\end{figure}
\begin{figure}
\begin{center}
	\caption{Schematic illustration of the outflow and the \desys\ of \iras\ Source A. 
			The \sam\ ($j$) of the gas in each structure 
			is shown in the unit of ($10^{-4}$ \kmps\ pc) (Table \ref{tb:sam}). 
			Two possibilities for the origin of the NW-SE outflow are presented. 
			(1) The NW-SE outflow structure originates from the protostar A1 and its circumstellar disk, 
			or (2) it is a disk wind from the circummultiple structure of Source A. 
			{\bff In this figure, we employ the distance ($d$) of 137 pc \citep{Ortiz-Leon_Ophiuchus} to \iras\ from the Sun. 			
			The distance between the protostars A1 and A2 is $\sim$50 au on the \pos\ \citep[e.g.][; \citetalias{Oya_16293Cyc4}]{HernandezGomez2019_contVLA}. 
			Its deprojected distance is 95.6 au \citep{Maureira2020} by employing 
			{\bf the \ia\ of 64\degr\ \citep{Maureira2020} and}
			$d$ of 141 pc \citep{Dzib2018}, 
			which corresponds to $\sim$93 au for $d$ of 137 pc. 
			The lower and upper limits to the radius of the circumstellar disk of A1 are 30 au and 50 au, respectively, 
			according to the \TFA\ observations by \citetalias{Oya_16293Cyc4}. 
			The existence of a circumstellar disk around A2 is still controversial; 
			\citet{Maureira2020} attributed the \TFA\ emission to a possible disk around A2 with a radius of 50 au using $d$ of 141 pc, 
			while \citetalias{Oya_16293Cyc4} attributed it to a skirt of the disk around A1. 
			The circummultiple structure is traced up to 300 au by the \CO\ emission reported by \citetalias{Oya_16293Cyc4}.} 
			\label{fig:scheme_sam}}
\end{center}
\end{figure}

\begin{figure}
\begin{center}
	\iffigure
	\epsscale{0.6}
	\plotone{\dirfig 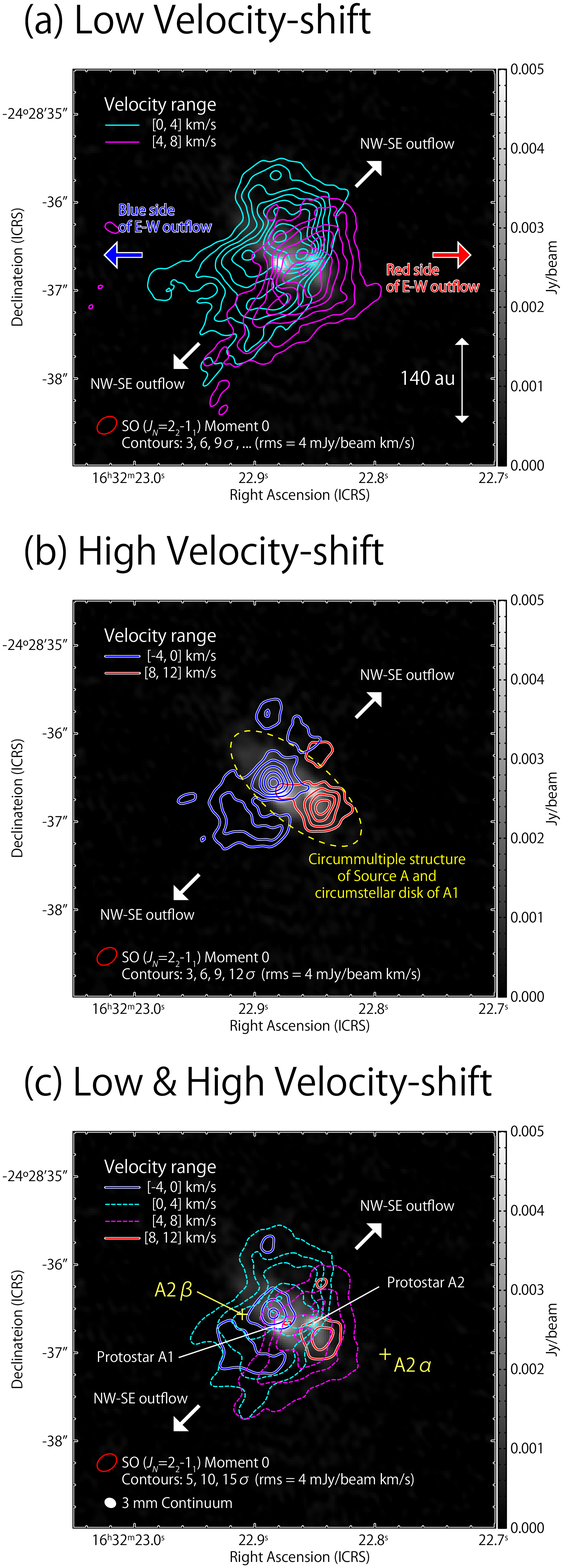} 
	\fi
\end{center}
\end{figure}
\begin{figure}
\begin{center}
	\caption{Integrated intensity maps of the SO (\so) line (contours): 
			the low velocity-shift components in panel (a), 
			the high velocity-shift components in panel (b), 
			and the overlap of them in panel (c). 
			Gray-scale maps show the 3.1 mm continuum image. 
			The velocity ranges for integration are shown in the top left corner of each panel: 
			from $-4$ to 0 \kmps\ for the blue contours, 
			from 0 to 4 \kmps\ for the cyan contours, 
			from 4 to 8 \kmps\ for the magenta contours, 
			and from 8 to 12 \kmps\ for the red contours. 
			The \sysV\ of Source A is \Vsys. 
			The cyan (0--4 \kmps) and magenta (4--8 \kmps) contours are 
			shown in dotted line in panel (c) for clarity. 
			Contour levels are every 3$\sigma$ in panels (a) and (b), 
			and every 5$\sigma$ in panel (c), 
			where the rms noise level is 4 \mJypb\ \kmps. 
			In panel (a), 
			the directions of the outflows on a $10^3$ au scale 
			reported previously are shown by four arrows: 
			the E-W bipolar outflow is reported by \citet{Yeh2008} and \citet{Loinard2013}, 
			the NW-SE one by \citet{vanderWiel2019}, 
			and both by \citet{Rao2009}, \citet{Kristensen2013}, and \citet{Girart2014}. 
			(See Section \ref{sec:intro_16293outflow}). 
			\label{fig:mom0_4kmps}}
\end{center}
\end{figure}

\begin{figure}
\begin{center}
	\iffigure
	\epsscale{0.6}
	\plotone{\dirfig 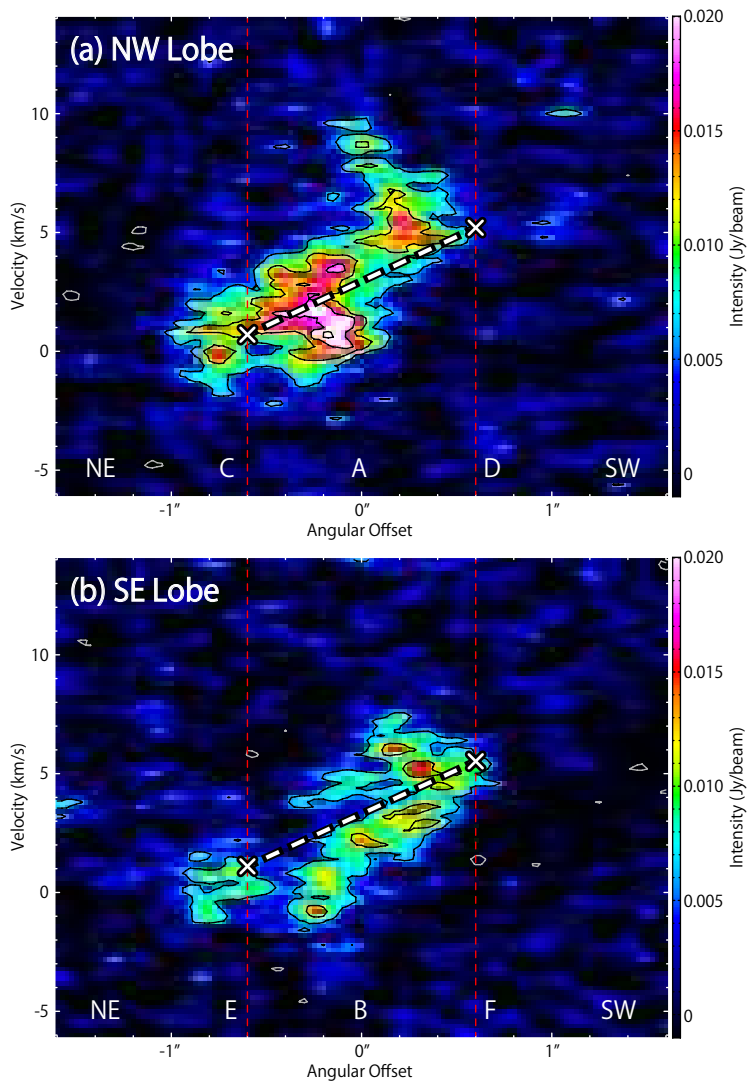}
	\fi
	\caption{\bff Position-velocity diagrams of the SO (\so) emission. 
			Their position axes are represented by red arrows in Figure \ref{fig:spectra}(a),  
			which are taken across the outflow axis. 
			The central positions of the position axes are taken at Positions A and B shown in Figure \ref{fig:spectra}. 
			Positions C, D, E, and F are represented by red dashed vertical lines in the diagrams. 
			White crosses in the diagrams represent 
			the velocity centroids obtained by the Gaussian fitting  for the spectra at Positions C, D, E, and F (Table \ref{tb:gaussfit_spectra}; Figure \ref{fig:spectra}). 
			The white dashed lines have their ends at the white crosses; 
			they show the overall trend of the velocity gradient across the outflow axis. 
			\label{fig:PV}}
\end{center}
\end{figure}

\begin{landscape}
\begin{figure}
\vspace*{-20pt}
\begin{center}
	\iffigure
	\epsscale{1.0}
	\plotone{\dirfig 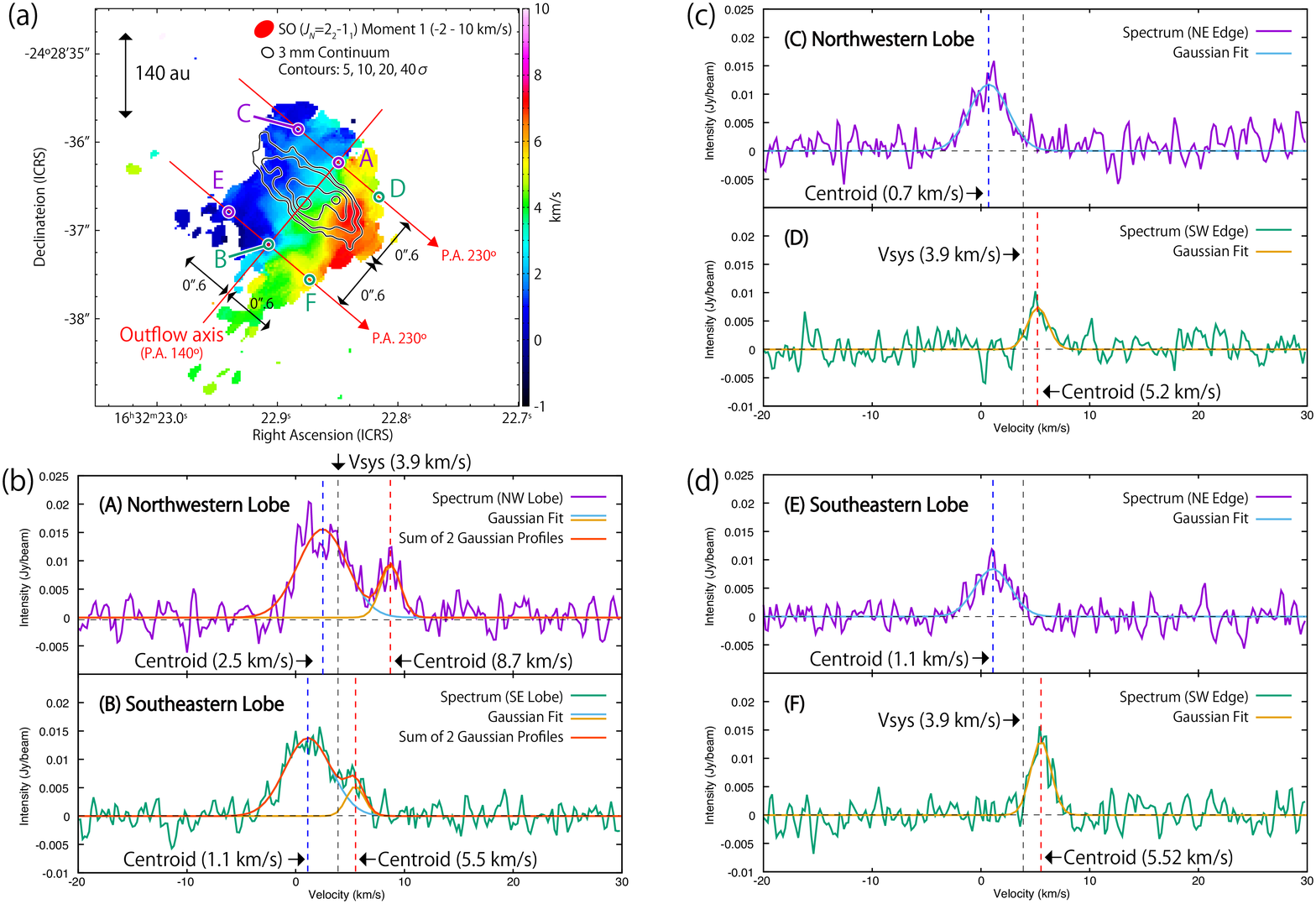} 
	\vspace*{-30pt}
	\fi
\end{center}
\end{figure}
\end{landscape}
\begin{landscape}
\begin{figure}
\begin{center}
	\caption{(a) Velocity (moment 1) map of the SO (\so) line. 
			The color map and the contours are the same as those in Figure \ref{fig:mom1}(b). 
			Six circles represent the positions where the spectra of the SO (\so) line in panels (b), (c), and (d) are taken. 
			Positions A, C, and D are on the northwestern outflow lobe, 
			while positions B, E, and F are on the southeastern lobe. 
			Position A is taken on the outflow axis at the distance of 0\farcs6 (80 au) from the protostar A1 on its northwestern side. 
			Positions C and D are taken at the northeastern and southwestern edges of the northwestern lobe, respectively, 
			which are at the distance of 0\farcs6 from position A. 
			Positions B, E, and F are as well, for the southeastern lobe. 
			(b) Spectra on the outflow axis taken at positions A (upper panel) and B (lower panel). 
			(c) Spectra in the northwestern lobe taken at positions C (upper panel) and D (lower panel). 
			(d) Spectra in the southeastern lobe taken at positions E (upper panel) and F (lower panel). 			
			The result of Gaussian fitting for each spectrum is shown in each panel 
			and summarized in Table \ref{tb:gaussfit_spectra}. 
			Spectra at positions A and B are fitted by two Gaussian profiles, 
			while the others are by one Gaussian profile. 
			\label{fig:spectra}}
\end{center}
\end{figure}
\end{landscape}

\appendix
\setcounter{table}{0}
\setcounter{figure}{0}
\makeatletter
	\def\thetable{A\the\c@table}
	\def\thefigure{A\the\c@figure}
\makeatother

\begin{table}
\begin{center}
	\caption{\bff Results of the Gaussian Fitting for the Spectra of the OCS (\ocsa) Line\tablenotemark{a}
			\label{tb:gaussfit_spectra_OCS}}	
	\vspace*{10pt}
	\begin{tabular}{lcccc}
	\hline \vspace*{-5pt} 
	Position & Distance from & Peak Intensity & Velocity Centroid & FWHM\tablenotemark{b}  \\ \vspace*{3pt} 
	& the Outflow Axis & (\mJypb) & (\kmps) & (\kmps)  \\ \hline \hline 
	\multicolumn{5}{c}{Northwestern Outflow Lobe} \\ \hline 
	Center (Position A) & 0\arcsec & $12.7 \pm 0.9$ & $3.6 \pm 0.1$ & $3.4 \pm 0.3$ \\ 
	& & $2.9 \pm 1.2$ & $8.9 \pm 0.4$ & $1.9 \pm 0.9$ \\ 
	NE Edge (Position C) & 0\farcs6 (80 au) & $20.6 \pm 1.4$ & $-1.28 \pm 0.06$ & $1.6 \pm 0.1$ \\ 
	& & $10.8 \pm 1.1$ & $3.0 \pm 0.1$ & $2.6 \pm 0.4$ \\ 
	\phantom{NE Edge} (Position C\arcmin) & 1\farcs0 (137 au) & $12.6 \pm 0.9$ & $3.7 \pm 0.1$ & $3.0 \pm 0.3$ \\ 
	SW Edge (Position D) & 0\farcs6 (80 au) & $8.4 \pm 1.2$ & $5.3 \pm 0.1$ & $1.8 \pm 0.3$ \\ 
	\phantom{SW Edge} (Position D\arcmin) & 1\farcs0 (137 au) & $9.5 \pm 1.7$ & $5.26 \pm 0.09$ & $1.0 \pm 0.2$ \\ 
	\hline 
	\multicolumn{5}{c}{Southeastern Outflow Lobe} \\ \hline 
	Center (Position B) & 0\arcsec & $5.3 \pm 0.9$ & $2.5 \pm 0.5$ & $3.1 \pm 1.1$ \\ 
	& & $9.4 \pm 2.3$ & $4.5 \pm 0.9$ & $1.1 \pm 0.3$ \\ 
	NE Edge (Position E) & 0\farcs6 (80 au) & $11.0 \pm 1.3$ & $0.8 \pm 0.1$ & $1.8 \pm 0.2$ \\ 
	& & $5.9 \pm 2.2$ & $4.0 \pm 0.1$ & $0.6 \pm 0.3$ \\ 
	\phantom{NE Edge} (Position E\arcmin) & 1\farcs0 (137 au) & $4.5 \pm 1.5$ & $2.9 \pm 0.2$ & $1.2 \pm 0.5$ \\ 
	SW Edge (Position F) & 0\farcs6 (80 au) & $17.7 \pm 1.1$ & $5.40 \pm 0.06$ & $1.9 \pm 0.1$ \\ 
	\phantom{SW Edge} (Position F\arcmin) & 1\farcs0 (137 au) & $14.7 \pm 1.1$ & $5.78 \pm 0.08$ & $2.0 \pm 0.2$ \\ 
	\hline
	\end{tabular}
	\tablenotetext{a}{\bff To derive the line parameters, each spectrum is taken
					at each position
					depicted in Figure \ref{fig:spectra_OCS}(a).
					\distanceRem  
					} 
	\tablenotetext{b}{\bff Full width at half maximum of the Gaussian profile.} 
\end{center}
\end{table}

\begin{landscape}
\begin{table}
\begin{center}
	\caption{\bff Specific Angular Momenta of the Outflow Lobes Traced by the OCS (\ocsa) Line\tablenotemark{a} 
			\label{tb:sam_OCS}}
	\begin{tabular}{lccccccc}
	\hline 
	Structure & Distance from & \multicolumn{6}{c}{Inclination Angle\tablenotemark{b}} \\ 
	& the Outflow Axis & 40\degr & 50\degr & 60\degr & 70\degr & 80\degr & 90\degr \\ \hline \hline
	NW Outflow Lobe & 80 au\tablenotemark{c} & $15.8 \pm 0.4$ & $13.3 \pm 0.3$ & $11.8 \pm 0.3$ & $10.8 \pm 0.2$ & $10.3 \pm 0.2$ & $10.2 \pm 0.2$ \\ 
	& 137 au & $8.1 \pm 0.7$ & $6.8 \pm 0.6$ & $6.0 \pm 0.5$ & $5.5 \pm 0.5$ & $5.3 \pm 0.5$ & $5.2 \pm 0.4$ \\ 
	SE Outflow Lobe & 80 au\tablenotemark{c} & $10.8 \pm 0.3$ & $9.1 \pm 0.2$ & $8.0 \pm 0.2$ & $7.4 \pm 0.2$ & $7.0 \pm 0.2$ & $6.9 \pm 0.2$ \\ 
	& 137 au & $14.9 \pm 1.1$ & $12.5 \pm 0.9$ & $11.0 \pm 0.8$ & $10.2 \pm 0.8$ & $9.7 \pm 0.7$ & $9.6 \pm 0.7$ \\ 
	\hline 
	\end{tabular}

\vspace*{-30pt}

	\tablenotetext{a}{\bff The \sams\ are in the unit of $10^{-4}$ \kmps\ pc. 
					The values in the unit of \kmps\ au can be obtained by 
					multiplying those in \kmps\ pc by $2.06 \times 10^5$. 
					The ranges of the \sams\ for the outflow lobes are 
					based on the uncertainties calculated by the propagation from the fitting errors of the velocity centroids (Table \ref{tb:gaussfit_spectra}). 
					\distanceRem 
					} 
	\tablenotetext{b}{\bff \incRemOutflow.}
	\tablenotetext{c}{\bff Spectra of the OCS (\ocsa) line at Positions C and E show double-Gaussian profiles.
					In the calculation of the \sam, 
					the weighted average of the two velocity centroids derived from the double Gaussian fitting is used (See Appendix \ref{sec:app_OCS}). 
					}
\end{center}
\end{table}
\end{landscape}

\begin{landscape}
\begin{figure}
\begin{center}
	\iffigure
	\epsscale{1.0}
	\plotone{\dirfig 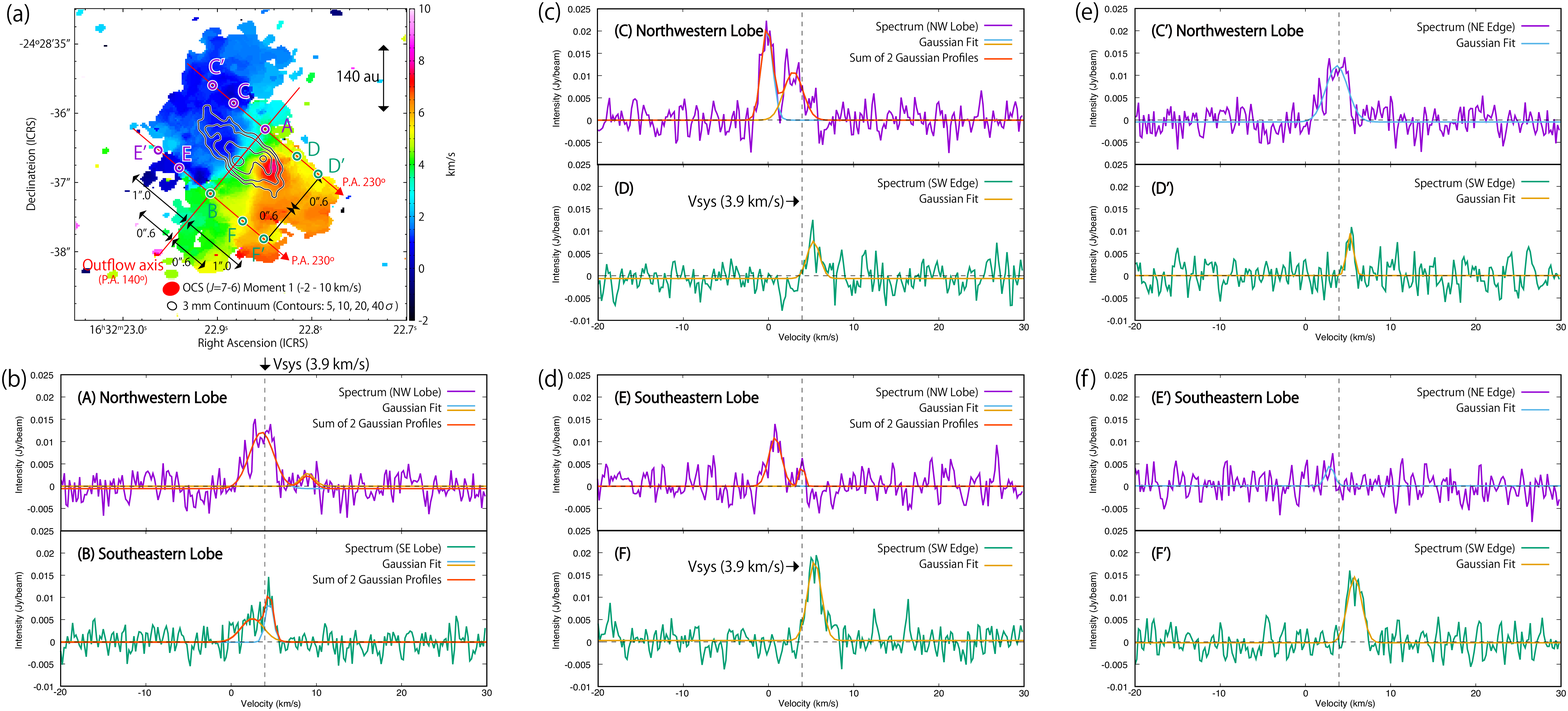}
	\fi
	\caption{\bff Same as Figure \ref{fig:spectra} for the OCS (\ocsa) line. 
			Four positions (C\arcmin, D\arcmin, E\arcmin, F\arcmin) are 
			taken at the distance of 1\farcs0 (137 au) from Position A or B, 
			as shown in panel (a). 
			\label{fig:spectra_OCS}}
\end{center}
\end{figure}
\end{landscape}


\begin{thebibliography}{}
\bibitem[ALMA Partnership(2017)]{ALMA_TH-C5} ALMA Partnership, 2017, S. Asayama, A. Biggs, I. de Gregorio, B. Dent, J. Di Francesco, E. Fomalont,A. Hales, J. Hibbard, G. Marconi, S. Kameno, B. Vila Vilaro, E. Villard, F. Stoehr, ISBN: 978-3-923524-66-2
\bibitem[Alves et al.(2017)]{Alves2017} Alves, F.~O., Girart, J.~M., Caselli, P., et al.\ 2017, \aap, 603, L3. doi:10.1051/0004-6361/201731077
\bibitem[Alves et al.(2019)]{Alves2019} Alves, F.~O., Caselli, P., Girart, J.~M., et al.\ 2019, Science, 366, 90
\bibitem[Anderson et al.(2003)]{Anderson2003} Anderson, J.~M., Li, Z.-Y., Krasnopolsky, R., et al.\ 2003, \apjl, 590, L107. doi:10.1086/376824
\bibitem[Bjerkeli et al.(2016)]{Bjerkeli2016c} Bjerkeli, P., van der Wiel, M.~H.~D., Harsono, D., et al.\ 2016, \nat, 540, 406
\bibitem[Bottinelli et al.(2004)]{Bottinelli2004} Bottinelli, S., Ceccarelli, C., Neri, R., et al.\ 2004, \apjl, 617, L69
\bibitem[Brinch et al.(2016)]{Brinch2016} Brinch, C., J{\o}rgensen, J.~K., Hogerheijde, M.~R., et al.\ 2016, \apjl, 830, L16. doi:10.3847/2041-8205/830/1/L16
\bibitem[Castets et al.(2001)]{Castets2001} Castets, A., Ceccarelli, C., Loinard, L., et al.\ 2001, \aap, 375, 40. doi:10.1051/0004-6361:20010662
\bibitem[Caux et al.(2011)]{Caux2011} Caux, E., Kahane, C., Castets, A., et al.\ 2011, \aap, 532, A23 
\bibitem[Cazaux et al.(2003)]{Cazaux2003} Cazaux, S., Tielens, A.~G.~G.~M., Ceccarelli, C., et al.\ 2003, \apjl, 593, L51
\bibitem[Ceccarelli(2004)]{Ceccarelli2004} Ceccarelli, C.\ 2004, Star Formation in the Interstellar Medium: In Honor of David Hollenbach, 323, 195 
\bibitem[Chandler et al.(2005)]{Chandler2005} Chandler, C.~J., Brogan, C.~L., Shirley, Y.~L., et al.\ 2005, \apj, 632, 371
\bibitem[Chen et al.(2013)]{Chen2013} Chen, X., Arce, H.~G., Zhang, Q., et al.\ 2013, \apj, 768, 110. doi:10.1088/0004-637X/768/2/110
\bibitem[Duch{\^e}ne \& Kraus(2013)]{Duchene2013} Duch{\^e}ne, G. \& Kraus, A.\ 2013, \araa, 51, 269. doi:10.1146/annurev-astro-081710-102602
\bibitem[Dzib et al.(2018)]{Dzib2018} Dzib, S.~A., Ortiz-Le{\'o}n, G.~N., Hern{\'a}ndez-G{\'o}mez, A., et al.\ 2018, \aap, 614, A20. doi:10.1051/0004-6361/201732093
\bibitem[Favre et al.(2014)]{Favre_SMA} Favre, C., J{\o}rgensen, J.~K., Field, D., et al.\ 2014, \apj, 790, 55 
\bibitem[Girart et al.(2014)]{Girart2014} Girart, J.~M., Estalella, R., Palau, A., et al.\ 2014, \apjl, 780, L11
\bibitem[Green \& Chapman(1978)]{Green1978} Green, S. \& Chapman, S.\ 1978, \apjs, 37, 169
\bibitem[Hara et al.(2020)]{Hara2020} Hara, C., Kawabe, R., Nakamura, F., et al.\ 2020, arXiv:2010.06825
\bibitem[Heath \& Nixon(2020)]{Heath2020} Heath, R.~M. \& Nixon, C.~J.\ 2020, \aap, 641, A64. doi:10.1051/0004-6361/202038548
\bibitem[Hern{\'a}ndez-G{\'o}mez et al.(2019)]{HernandezGomez2019_contVLA} Hern{\'a}ndez-G{\'o}mez, A., Loinard, L., Chandler, C.~J., et al.\ 2019, \apj, 875, 94
\bibitem[Hirano et al.(2001)]{Hirano2001} Hirano, N., Mikami, H., Umemoto, T., et al.\ 2001, \apj, 547, 899. doi:10.1086/318432
\bibitem[Hirota et al.(2017)]{Hirota2017} Hirota, T., Machida, M.~N., Matsushita, Y., et al.\ 2017, Nature Astronomy, 1, 0146
\bibitem[Huang et al.(2005)]{Huang2005} Huang, H.-C., Kuan, Y.-J., Charnley, S.~B., et al.\ 2005, Advances in Space Research, 36, 146. doi:10.1016/j.asr.2005.03.115
\bibitem[J{\o}rgensen et al.(2011)]{Jorgensen2011} J{\o}rgensen, J.~K., Bourke, T.~L., Nguyen Luong, Q., et al.\ 2011, \aap, 534, A100. doi:10.1051/0004-6361/201117139
\bibitem[J{\o}rgensen et al.(2012)]{Jorgensen_sugar} J{\o}rgensen, J.~K., Favre, C., Bisschop, S.~E., et al.\ 2012, \apjl, 757, L4 
\bibitem[J{\o}rgensen et al.(2016)]{Jorgensen_PILS} J{\o}rgensen, J.~K., van der Wiel, M.~H.~D., Coutens, A., et al.\ 2016, \aap, 595, A117
\bibitem[Kristensen et al.(2013)]{Kristensen2013} Kristensen, L.~E., Klaassen, P.~D., Mottram, J.~C., et al.\ 2013, \aap, 549, L6
\bibitem[Kuruwita et al.(2017)]{Kuruwita2017} Kuruwita, R.~L., Federrath, C., \& Ireland, M.\ 2017, \mnras, 470, 1626. doi:10.1093/mnras/stx1299
\bibitem[Kuan et al.(2004)]{Kuan2004} Kuan, Y.-J., Huang, H.-C., Charnley, S.~B., et al.\ 2004, \apjl, 616, L27
\bibitem[Lee et al.(2021)]{Lee2021} Lee, C.-F., Tabone, B., Cabrit, S., et al.\ 2021, \apjl, 907, L41. doi:10.3847/2041-8213/abda38
\bibitem[Lique et al.(2006a)]{Lique2006a} Lique, F., Dubernet, M.-L., Spielfiedel, A., et al.\ 2006, \aap, 450, 399
\bibitem[Lique et al.(2006b)]{Lique2006b} Lique, F., Spielfiedel, A., \& Cernicharo, J.\ 2006, \aap, 451, 1125. doi:10.1051/0004-6361:20054363
\bibitem[Loinard et al.(2013)]{Loinard2013} Loinard, L., Zapata, L.~A., Rodriguez, L.~F., et al.\ 2013, \mnras, 430, L10
\bibitem[Looney et al.(2000)]{Looney2000} Looney, L.~W., Mundy, L.~G., \& Welch, W.~J.\ 2000, \apj, 529, 477 
\bibitem[Lykke et al.(2017)]{Lykke2017} Lykke, J.~M., Coutens, A., J{\o}rgensen, J.~K., et al.\ 2017, \aap, 597, A53. doi:10.1051/0004-6361/201629180
\bibitem[Machida et al.(2009)]{Machida2009} Machida, M.~N., Inutsuka, S., \& Matsumoto, T.\ 2009, \apjl, 704, L10. doi:10.1088/0004-637X/704/1/L10
\bibitem[Maureira et al.(2020)]{Maureira2020} Maureira, M.~J., Pineda, J.~E., Segura-Cox, D.~M., et al.\ 2020, \apj, 897, 59
\bibitem[McMullin et al.(2007)]{McMullin_CASA} McMullin, J.~P., Waters, B., Schiebel, D., et al.\ 2007, Astronomical Data Analysis Software and Systems XVI, 376, 127
\bibitem[Miranda et al.(2017)]{Miranda2017} Miranda, R., Mu{\~n}oz, D.~J., \& Lai, D.\ 2017, \mnras, 466, 1170. doi:10.1093/mnras/stw3189
\bibitem[Mizuno et al.(1990)]{Mizuno1990} Mizuno, A., Fukui, Y., Iwata, T., et al.\ 1990, \apj, 356, 184. doi:10.1086/168829
\bibitem[Moody et al.(2019)]{Moody2019} Moody, M.~S.~L., Shi, J.-M., \& Stone, J.~M.\ 2019, \apj, 875, 66. doi:10.3847/1538-4357/ab09ee
\bibitem[M{\"u}ller et al.(2005)]{Muller_CDMS} M{\"u}ller, H.~S.~P., Schl{\"o}der, F., Stutzki, J., \& Winnewisser, G.\ 2005, Journal of Molecular Structure, 742, 215 
\bibitem[Mundy et al.(1992)]{Mundy1992} Mundy, L.~G., Wootten, A., Wilking, B.~A., et al.\ 1992, \apj, 385, 306. doi:10.1086/170939
\bibitem[Ortiz-Le{\'o}n et al.(2017)]{Ortiz-Leon_Ophiuchus} Ortiz-Le{\'o}n, G.~N., Loinard, L., Kounkel, M.~A., et al.\ 2017, \apj, 834, 141 
\bibitem[Oya et al.(2014)]{Oya_15398} Oya, Y., Sakai, N., Sakai, T., et al.\ 2014, \apj, 795, 152
\bibitem[Oya et al.(2016)]{Oya_16293} Oya, Y., Sakai, N., L{\'o}pez-Sepulcre, A., et al.\ 2016, \apj, 824, 88 
\bibitem[Oya et al.(2018)]{Oya_16293B} Oya, Y., Moriwaki, K., Onishi, S., et al.\ 2018, \apj, 854, 96
\bibitem[Oya et al.(2018)]{Oya_483outflow} Oya, Y., Sakai, N., Watanabe, Y., et al.\ 2018, \apj, 863, 72. doi:10.3847/1538-4357/aacf42
\bibitem[Oya \& Yamamoto(2020)]{Oya_16293Cyc4} Oya, Y. \& Yamamoto, S.\ 2020, \apj, 904, 185. doi:10.3847/1538-4357/abbe14
\bibitem[Pech et al.(2010)]{Pech2010} Pech, G., Loinard, L., Chandler, C.~J., et al.\ 2010, \apj, 712, 1403
\bibitem[Pickett et al.(1998)]{Pickett_JPL} Pickett, H.~M., Poynter, R.~L., Cohen, E.~A., et al.\ 1998, \jqsrt, 60, 883
\bibitem[Pineda et al.(2012)]{Pineda_ALMA} Pineda, J.~E., Maury, A.~J., Fuller, G.~A., et al.\ 2012, \aap, 544, L7 
\bibitem[Rao et al.(2009)]{Rao2009} Rao, R., Girart, J.~M., Marrone, D.~P., et al.\ 2009, \apj, 707, 921
\bibitem[Saiki \& Machida(2020)]{Saiki2020} Saiki, Y. \& Machida, M.~N.\ 2020, \apjl, 897, L22
\bibitem[Santangelo et al.(2015)]{Santangelo2015} Santangelo, G., Codella, C., Cabrit, S., et al.\ 2015, \aap, 584, A126. doi:10.1051/0004-6361/201526323
\bibitem[Sch{\"o}ier et al.(2002)]{Schoier_hotcore} Sch{\"o}ier, F.~L., J{\o}rgensen, J.~K., van Dishoeck, E.~F., \& Blake, G.~A.\ 2002
\bibitem[Sch{\"o}ier et al.(2005)]{Schoier_LAMDA} Sch{\"o}ier, F.~L., van der Tak, F.~F.~S., van Dishoeck, E.~F., et al.\ 2005, \aap, 432, 369
\bibitem[Stark et al.(2004)]{Stark2004} Stark, R., Sandell, G., Beck, S.~C., et al.\ 2004, \apj, 608, 341
\bibitem[Tabone et al.(2017)]{Tabone2017} Tabone, B., Cabrit, S., Bianchi, E., et al.\ 2017, \aap, 607, L6. doi:10.1051/0004-6361/201731691
\bibitem[Tabone et al.(2020)]{Tabone2020} Tabone, B., Cabrit, S., Pineau des For{\^e}ts, G., et al.\ 2020, \aap, 640, A82. doi:10.1051/0004-6361/201834377
\bibitem[Tiede et al.(2020)]{Tiede2020} Tiede, C., Zrake, J., MacFadyen, A., et al.\ 2020, \apj, 900, 43. doi:10.3847/1538-4357/aba432
\bibitem[Tobin et al.(2015)]{Tobin2015} Tobin, J.~J., Looney, L.~W., Wilner, D.~J., et al.\ 2015, \apj, 805, 125. doi:10.1088/0004-637X/805/2/125
\bibitem[Tobin et al.(2016)]{Tobin2016} Tobin, J.~J., Kratter, K.~M., Persson, M.~V., et al.\ 2016, \nat, 538, 483
\bibitem[van der Wiel et al.(2019)]{vanderWiel2019} van der Wiel, M.~H.~D., Jacobsen, S.~K., J{\o}rgensen, J.~K., et al.\ 2019, \aap, 626, A93
\bibitem[van Dishoeck et al.(1995)]{vanDishoeck1995} van Dishoeck, E.~F., Blake, G.~A., Jansen, D.~J., et al.\ 1995, \apj, 447, 760
\bibitem[Wakelam et al.(2004)]{Wakelam2004} Wakelam, V., Castets, A., Ceccarelli, C., et al.\ 2004, \aap, 413, 609. doi:10.1051/0004-6361:20031572
\bibitem[Walker et al.(1988)]{Walker1988} Walker, C.~K., Lada, C.~J., Young, E.~T., et al.\ 1988, \apj, 332, 335. doi:10.1086/166659
\bibitem[Watanabe et al.(2017)]{Watanabe2017} Watanabe, Y., Sakai, N., L{\'o}pez-Sepulcre, A., et al.\ 2017, \apj, 847, 108. doi:10.3847/1538-4357/aa88b6
\bibitem[Wootten(1989)]{Wootten1989} Wootten, A.\ 1989, \apj, 337, 858
\bibitem[Yamamoto(2017)]{Yamamoto2017} Yamamoto, S.\ 2017, Introduction to Astrochemistry: Chemical Evolution from Interstellar Clouds to Star and Planet Formation, Astronomy and Astrophysics Library, by Satoshi Yamamoto. ISBN 978-4-431-54170-7. Springer Japan, 2017
\bibitem[Yang et al.(2010)]{Yang2010} Yang, B., Stancil, P.~C., Balakrishnan, N., et al.\ 2010, \apj, 718, 1062
\bibitem[Yeh et al.(2008)]{Yeh2008} Yeh, S.~C.~C., Hirano, N., Bourke, T.~L., et al.\ 2008, \apj, 675, 454
\bibitem[Zhang et al.(2018)]{Zhang2018} Zhang, Y., Higuchi, A.~E., Sakai, N., et al.\ 2018, \apj, 864, 76
\end{thebibliography}

\begin{thebibliography}{}
\bibitem[ALMA Partnership(2016)]{ALMA_TH-C4} ALMA Partnership, 2016, S. Asayama, A. Biggs, I. de Gregorio, B. Dent, J. Di Francesco, E. Fomalont,A. Hales, E. Humphries, S. Kameno, E. M\"{u}ller, B. Vila Vilaro, E. Villard, F. Stoehr, ``ALMA Cycle 4 Technical Handbook'', ISBN 978-3-923524-66-2
\bibitem[Aikawa et al.(1996)]{Aikawa1996} Aikawa, Y., Miyama, S.~M., Nakano, T., et al.\ 1996, \apj, 467, 684
\bibitem[Alves et al.(2019)]{Alves2019} Alves, F.~O., Caselli, P., Girart, J.~M., et al.\ 2019, Science, 366, 90
\bibitem[Artur de la Villarmois et al.(2018)]{ArturdelaVillarmois2018} Artur de la Villarmois, E., Kristensen, L.~E., J{\o}rgensen, J.~K., et al.\ 2018, \aap, 614, A26
\bibitem[Aso et al.(2017)]{Aso2017} Aso, Y., Ohashi, N., Aikawa, Y., et al.\ 2017, \apj, 849, 56
\bibitem[Bate \& Bonnell(1997)]{Bate1997} Bate, M.~R., \& Bonnell, I.~A.\ 1997, \mnras, 285, 33
\bibitem[Bate et al.(2003)]{Bate2003} Bate, M.~R., Bonnell, I.~A., \& Bromm, V.\ 2003, \mnras, 339, 577
\bibitem[Boehler et al.(2017)]{Boehler2017} Boehler, Y., Weaver, E., Isella, A., et al.\ 2017, \apj, 840, 60
\bibitem[Bottinelli et al.(2004)]{Bottinelli2004} Bottinelli, S., Ceccarelli, C., Neri, R., et al.\ 2004, \apjl, 617, L69
\bibitem[Caux et al.(2011)]{Caux2011} Caux, E., Kahane, C., Castets, A., et al.\ 2011, \aap, 532, A23 
\bibitem[Cazaux et al.(2003)]{Cazaux2003} Cazaux, S., Tielens, A.~G.~G.~M., Ceccarelli, C., et al.\ 2003, \apjl, 593, L51
\bibitem[Ceccarelli(2004)]{Ceccarelli2004} Ceccarelli, C.\ 2004, Star Formation in the Interstellar Medium: In Honor of David Hollenbach, 323, 195 
\bibitem[Chandler et al.(2005)]{Chandler2005} Chandler, C.~J., Brogan, C.~L., Shirley, Y.~L., et al.\ 2005, \apj, 632, 371
{\bff \bibitem[Chen et al.(2013)]{Chen2013} Chen, X., Arce, H.~G., Zhang, Q., et al.\ 2013, \apj, 768, 110} 
\bibitem[Coutens et al.(2016)]{Coutens2016} Coutens, A., J{\o}rgensen, J.~K., van der Wiel, M.~H.~D., et al.\ 2016, \aap, 590, L6
\bibitem[Dutrey et al.(2014)]{Dutrey2014} Dutrey, A., di Folco, E., Guilloteau, S., et al.\ 2014, \nat, 514, 600
\bibitem[Endres et al.(2016)]{Endres_CDMS} Endres, C.~P., Schlemmer, S., Schilke, P., et al.\ 2016, Journal of Molecular Spectroscopy, 327, 95
\bibitem[Fateeva et al.(2011)]{Fateeva2011} Fateeva, A.~M., Bisikalo, D.~V., Kaygorodov, P.~V., et al.\ 2011, \apss, 335, 125
\bibitem[Favre et al.(2014)]{Favre_SMA} Favre, C., J{\o}rgensen, J.~K., Field, D., et al.\ 2014, \apj, 790, 55 
\bibitem[Harsono et al.(2018)]{Harsono2018} Harsono, D., Bjerkeli, P., van der Wiel, M.~H.~D., et al.\ 2018, Nature Astronomy, 2, 646
\bibitem[Hartmann(2009)]{Hartmann2009} Hartmann, L.\ 2009, Accretion Processes in Star Formation: Second Edition
\bibitem[Hennebelle, \& Ciardi(2009)]{Hennebelle2009} Hennebelle, P., \& Ciardi, A.\ 2009, \aap, 506, L29
\bibitem[Hern{\'a}ndez-G{\'o}mez et al.(2019)]{HernandezGomez2019_contVLA} Hern{\'a}ndez-G{\'o}mez, A., Loinard, L., Chandler, C.~J., et al.\ 2019, \apj, 875, 94
\bibitem[Huang et al.(2005)]{Huang2005} Huang, H.-C., Kuan, Y.-J., Charnley, S.~B., et al.\ 2005, Advances in Space Research, 36, 146
\bibitem[Imai et al.(2016)]{Imai2016} Imai, M., Sakai, N., Oya, Y., et al.\ 2016, \apjl, 830, L37
\bibitem[Imai et al.(2019)]{Imai2019} Imai, M., Oya, Y., Sakai, N., et al.\ 2019, \apjl, 873, L21
\bibitem[Jacobsen et al.(2018)]{Jacobsen2018_16293temp} Jacobsen, S.~K., J{\o}rgensen, J.~K., van der Wiel, M.~H.~D., et al.\ 2018, \aap, 612, A72
\bibitem[Jacobsen et al.(2019)]{Jacobsen2019} Jacobsen, S.~K., J{\o}rgensen, J.~K., Di Francesco, J., et al.\ 2019, \aap, 629, A29
\bibitem[J{\o}rgensen et al.(2012)]{Jorgensen_sugar} J{\o}rgensen, J.~K., Favre, C., Bisschop, S.~E., et al.\ 2012, \apjl, 757, L4 
\bibitem[J{\o}rgensen et al.(2016)]{Jorgensen_PILS} J{\o}rgensen, J.~K., van der Wiel, M.~H.~D., Coutens, A., et al.\ 2016, \aap, 595, A117
\bibitem[Kratter et al.(2008)]{Kratter2008} Kratter, K.~M., Matzner, C.~D., \& Krumholz, M.~R.\ 2008, \apj, 681, 375
\bibitem[Kuan et al.(2004)]{Kuan2004} Kuan, Y.-J., Huang, H.-C., Charnley, S.~B., et al.\ 2004, \apjl, 616, L27
\bibitem[Lam et al.(2019)]{Lam2019} Lam, K.~H., Li, Z.-Y., Chen, C.-Y., et al.\ 2019, \mnras, 489, 5326
\bibitem[Lee et al.(2017)]{Lee2017_HH212} Lee, C.-F., Li, Z.-Y., Ho, P.~T.~P., et al.\ 2017, \apj, 843, 27
\bibitem[Li et al.(2011)]{Li2011} Li, Z.-Y., Krasnopolsky, R., \& Shang, H.\ 2011, \apj, 738, 180
\bibitem[Lim et al.(2016)]{Lim2016} Lim, J., Hanawa, T., Yeung, P.~K.~H., et al.\ 2016, \apj, 831, 90
\bibitem[Loinard et al.(2007)]{Loinard2007} Loinard, L., Chandler, C.~J., Rodr{\'\i}guez, L.~F., et al.\ 2007, \apj, 670, 1353
\bibitem[Looney et al.(2000)]{Looney2000} Looney, L.~W., Mundy, L.~G., \& Welch, W.~J.\ 2000, \apj, 529, 477 
\bibitem[Lykke et al.(2017)]{Lykke2017} Lykke, J.~M., Coutens, A., J{\o}rgensen, J.~K., et al.\ 2017, \aap, 597, A53
\bibitem[Machida et al.(2011)]{Machida2011} Machida, M.~N., Inutsuka, S.-I., \& Matsumoto, T.\ 2011, \pasj, 63, 555
\bibitem[Maeda et al.(2008)]{Maeda2008} Maeda, A., Medvedev, I.~R., Winnewisser, M., et al.\ 2008, \apjs, 176, 543
{\bff \bibitem[Manigand et al.(2020a)]{Manigand2020a} Manigand, S., J{\o}rgensen, J.~K., Calcutt, H., et al.\ 2020, \aap, 635, A48} 
{\bff \bibitem[Manigand et al.(2020b)]{Manigand2020b} Manigand, S., Coutens, A., Loison, J.-C., et al.\ 2020, arXiv:2007.04000} 
\bibitem[Matsumoto \& Hanawa(2003)]{Matsumoto2003} Matsumoto, T., \& Hanawa, T.\ 2003, \apj, 595, 913
\bibitem[Matsumoto et al.(2019)]{Matsumoto2019} Matsumoto, T., Saigo, K., \& Takakuwa, S.\ 2019, \apj, 871, 36
\bibitem[Maureira et al.(2020)]{Maureira2020} Maureira, M.~J., Pineda, J.~E., Segura-Cox, D.~M., et al.\ 2020, \apj, 897, 59
\bibitem[M{\"u}ller et al.(2005)]{Muller_CDMS} M{\"u}ller, H.~S.~P., Schl{\"o}der, F., Stutzki, J., \& Winnewisser, G.\ 2005, Journal of Molecular Structure, 742, 215 
\bibitem[M{\"u}ller et al.(2019)]{Muller2019} M{\"u}ller, H.~S.~P., Maeda, A., Thorwirth, S., et al.\ 2019, \aap, 621, A143
\bibitem[Murillo et al.(2013)]{Murillo2013} Murillo, N.~M., Lai, S.-P., Bruderer, S., et al.\ 2013, \aap, 560, A103
\bibitem[Ohashi et al.(2014)]{Ohashi2014} Ohashi, N., Saigo, K., Aso, Y., et al.\ 2014, \apj, 796, 131
\bibitem[Okoda et al.(2018)]{Okoda2018} Okoda, Y., Oya, Y., Sakai, N., et al.\ 2018, \apjl, 864, L25
\bibitem[Ortiz-Le{\'o}n et al.(2017)]{Ortiz-Leon_Ophiuchus} Ortiz-Le{\'o}n, G.~N., Loinard, L., Kounkel, M.~A., et al.\ 2017, \apj, 834, 141 
\bibitem[Oya et al.(2014)]{Oya_15398} Oya, Y., Sakai, N., Sakai, T., et al.\ 2014, \apj, 795, 152
\bibitem[Oya et al.(2016)]{Oya_16293} Oya, Y., Sakai, N., L{\'o}pez-Sepulcre, A., et al.\ 2016, \apj, 824, 88 
\bibitem[Oya et al.(2017)]{Oya_483} Oya, Y., Sakai, N., L{\'o}pez-Sepulcre, A., et al.\ 2017, \apj, 837, 174
{\bff \bibitem[Oya(2017)]{Oya_PhD} Oya, Y., ``A Few Tens au Scale Physical and Chemical Structures around Young Low-Mass Protostars'', The University of Tokyo, Japan} 
\bibitem[Oya et al.(2019)]{Oya_Elias29} Oya, Y., L{\'o}pez-Sepulcre, A., Sakai, N., et al.\ 2019, \apj, 881, 112
{\bff \bibitem[Oya and Yamamoto(in prep.)]{Oya_FERIA} Oya, Y., and Yamamoto, S. 2020, in preperation.} 
\bibitem[Pech et al.(2010)]{Pech2010} Pech, G., Loinard, L., Chandler, C.~J., et al.\ 2010, \apj, 712, 1403
\bibitem[Pickett et al.(1998)]{Pickett_JPL} Pickett, H.~M., Poynter, R.~L., Cohen, E.~A., et al.\ 1998, \jqsrt, 60, 883
\bibitem[Pineda et al.(2012)]{Pineda_ALMA} Pineda, J.~E., Maury, A.~J., Fuller, G.~A., et al.\ 2012, \aap, 544, L7 
\bibitem[Price et al.(2018)]{Price2018} Price, D.~J., Cuello, N., Pinte, C., et al.\ 2018, \mnras, 477, 1270
\bibitem[Ragusa et al.(2017)]{Ragusa2017} Ragusa, E., Dipierro, G., Lodato, G., et al.\ 2017, \mnras, 464, 1449
{\bff \bibitem[Sadavoy et al.(2018)]{Sadavoy2018} Sadavoy, S.~I., Myers, P.~C., Stephens, I.~W., et al.\ 2018, \apj, 869, 115} 
\bibitem[Sakai et al.(2014a)]{Sakai_1527nature} Sakai, N., Sakai, T., Hirota, T., et al.\ 2014a, \nat, 507, 78 
\bibitem[Sakai et al.(2014b)]{Sakai_1527apjl} Sakai, N., Oya, Y., Sakai, T., et al.\ 2014b, \apjl, 791, L38 
\bibitem[Satsuka et al.(2017)]{Satsuka2017} Satsuka, T., Tsuribe, T., Tanaka, S., et al.\ 2017, \mnras, 465, 986
\bibitem[Sch{\"o}ier et al.(2002)]{Schoier_hotcore} Sch{\"o}ier, F.~L., J{\o}rgensen, J.~K., van Dishoeck, E.~F., \& Blake, G.~A.\ 2002
\bibitem[Seifried et al.(2016)]{Seifried2016} Seifried, D., S{\'a}nchez-Monge, {\'A}., Walch, S., et al.\ 2016, \mnras, 459, 1892
\bibitem[Shi et al.(2012)]{Shi2012} Shi, J.-M., Krolik, J.~H., Lubow, S.~H., et al.\ 2012, \apj, 749, 118
\bibitem[Takakuwa et al.(2007)]{Takakuwa2007b} Takakuwa, S., Ohashi, N., Bourke, T.~L., et al.\ 2007b, \apj, 662, 431 
\bibitem[Takakuwa et al.(2014)]{Takakuwa2014} Takakuwa, S., Saito, M., Saigo, K., et al.\ 2014, \apj, 796, 1
\bibitem[Takakuwa et al.(2017)]{Takakuwa2017} Takakuwa, S., Saigo, K., Matsumoto, T., et al.\ 2017, \apj, 837, 86	
\bibitem[Tobin et al.(2015)]{Tobin_Perseus} Tobin, J.~J., Looney, L.~W., Wilner, D.~J., et al.\ 2015, \apj, 805, 125
{\bff \bibitem[Tobin et al.(2016a)]{Tobin2016a} Tobin, J.~J., Looney, L.~W., Li, Z.-Y., et al.\ 2016, \apj, 818, 73} 
\bibitem[Tobin et al.(2016b)]{Tobin2016b} Tobin, J.~J., Kratter, K.~M., Persson, M.~V., et al.\ 2016, \nat, 538, 483
\bibitem[Tokuda et al.(2014)]{Tokuda2014} Tokuda, K., Onishi, T., Saigo, K., et al.\ 2014, \apjl, 789, L4
\bibitem[Tomida et al.(2015)]{Tomida2015} Tomida, K., Okuzumi, S., \& Machida, M.~N.\ 2015, \apj, 801, 117
\bibitem[Tsukamoto et al.(2017)]{Tsukamoto2017} Tsukamoto, Y., Okuzumi, S., Iwasaki, K., et al.\ 2017, \pasj, 69, 95
\bibitem[van der Tak et al.(2007)]{vanderTak_radex} van der Tak, F.~F.~S., Black, J.~H., Sch{\"o}ier, F.~L., Jansen, D.~J., \& van Dishoeck, E.~F.\ 2007, \aap, 468, 627 
\bibitem[van der Wiel et al.(2019)]{vanderWiel2019} van der Wiel, M.~H.~D., Jacobsen, S.~K., J{\o}rgensen, J.~K., et al.\ 2019, \aap, 626, A93
\bibitem[van Dishoeck et al.(1995)]{vanDishoeck1995} van Dishoeck, E.~F., Blake, G.~A., Jansen, D.~J., et al.\ 1995, \apj, 447, 760
\bibitem[van't Hoff et al.(2020)]{vantHoff2020} van't Hoff, M.~L.~R., van Dishoeck, E.~F., J{\o}rgensen, J.~K., et al.\ 2020, \aap, 633, A7
\bibitem[Wootten(1989)]{Wootten1989} Wootten, A.\ 1989, \apj, 337, 858
\bibitem[Yeh et al.(2008)]{Yeh2008} Yeh, S.~C.~C., Hirano, N., Bourke, T.~L., et al.\ 2008, \apj, 675, 454
\bibitem[Yen et al.(2013)]{Yen2013} Yen, H.-W., Takakuwa, S., Ohashi, N., \& Ho, P.~T.~P.\ 2013, \apj, 772, 22 
\bibitem[Yen et al.(2017)]{Yen2017} Yen, H.-W., Koch, P.~M., Takakuwa, S., et al.\ 2017, \apj, 834, 178
\end{thebibliography}
\end{document}